\documentclass[rmp,aps,twocolumn,amsmath,amssymb,preprintnumbers]{revtex4}
\usepackage{natbib}
\usepackage{amsmath} \usepackage{amsfonts} \usepackage{amssymb}
\usepackage[utf8]{inputenc}
\usepackage{bbm}
\usepackage{epsfig}
\usepackage{psfrag}
\usepackage{subfigure}
\usepackage{graphics}
\usepackage{graphicx}
\usepackage{upgreek}
\renewcommand{\vec}[1]{\ensuremath{\mathbf{#1}}}

\newcommand{\dderiv}{\mathrm{d}}

\newenvironment{packed_enum}{
\begin{enumerate}
  \setlength{\itemsep}{1pt}
  \setlength{\parskip}{0pt}
  \setlength{\parsep}{0pt}
}{\end{enumerate}}

\newenvironment{packed_itemize}{
\begin{itemize}
  \setlength{\itemsep}{1pt}
  \setlength{\parskip}{0pt}
  \setlength{\parsep}{0pt}
}{\end{itemize}}

\begin{document}

\title{Too few spots in the cosmic microwave background}
\author{Youness Ayaita, Maik Weber, Christof Wetterich}
\affiliation{Institut f\"ur Theoretische Physik,
Universit\"at Heidelberg\\
Philosophenweg 16, D-69120 Heidelberg, Germany}

\begin{abstract}
	We investigate the abundance of large-scale hot and cold spots in
	the \mbox{WMAP-5} temperature maps and find considerable
	discrepancies compared to Gaussian simulations based on the
	$\Lambda$CDM best-fit model. Too few spots are present in the
	reliably observed cosmic microwave background (CMB) region, i.e., outside the
	foreground-contaminated parts excluded by the KQ75 mask. Even
	simulated maps created from the original \mbox{WMAP-5} estimated
	multipoles contain more spots than visible in the measured CMB
	maps. A strong suppression of the lowest multipoles would lead to better agreement. 
	The lack of spots is reflected in a low mean temperature
	fluctuation on scales of
	several degrees ($4^\circ$--$8^\circ$), which is only shared by less than
	$1\%$ ($0.16\%$--$0.62\%$)
	 of Gaussian $\Lambda$CDM simulations. After removing the quadrupole, the probabilities 
	change to $2.5\%$--$8.0\%$. This shows the importance of the anomalously low quadrupole for the statistical significance of the missing spots. We also analyze a possible
	violation of Gaussianity or statistical isotropy (spots are
	distributed differently outside and inside the masked region). 
\end{abstract}

\maketitle

\section{Introduction}

The precise measurement of anisotropies in the cosmic microwave
background (CMB) has played a key role in amplifying our knowledge
about the structure and evolution of the Universe. The best data
available today is provided by the Wilkinson Microwave Anisotropy
Probe (WMAP) satellite mission from five years of observation.  Its
results are powerful enough to put various cosmological models to
stringent tests. They helped establishing today's standard model of a
spatially flat universe with Gaussian initial perturbations, possibly
generated during an early inflationary epoch. According to the
standard $\Lambda$CDM model, the present Universe is essentially made
up from dark energy in the form of a cosmological constant $\Lambda$
and cold dark matter (CDM). Under the assumptions of Gaussianity and
statistical isotropy, all the information about the temperature
fluctuations in the CMB are encoded in the angular power spectrum
$C_\ell$ from a harmonic decomposition of the temperature field. A
crucial result of the WMAP analysis therefore is an estimate of the
multipoles $C_\ell$ which is in good agreement with the $\Lambda$CDM best
fit~\citep{Nolta09} except for the well-known discrepancies of the low
multipoles, especially the quadrupole $C_2$~\citep{Hinshaw07}.
Nonetheless, many issues are still under intense discussion.
Repeatedly, authors have claimed to detect non-Gaussian
signals~\citep{McEwen08,Yadav07} or statistical
anisotropy~\citep{Eriksen03, Hansen08, deOliveiraCosta03, Hoftuft09,
Land05, Bernui06}. Since the power spectrum is insensitive to these
anomalies, it is necessary to perform additional investigations of the
temperature sky map. These are done in harmonic, wavelet, and pixel
space~\citep{Cabella04}. Even if Gaussianity holds, it may still give
new insights to switch to another representation of the statistical
properties of the temperature maps since a phenomenon can be more
easily detected in one representation than in another.

The goal of this work is to provide a clear and intuitive analysis in
pixel space regarding abundances of large-scale hot and cold spots
identified as regions whose mean temperature contrasts exceed some
(variable) threshold. We analyze both observed CMB maps and Gaussian
simulations based on $\Lambda$CDM. The comparison reveals severe
deviations. Other authors who worked with statistics of local extrema
in the temperature field also observed significant
anomalies~\citep{Larson04, Larson05, Hou09}. 

We start by recalling some basic results that connect pixel-space
analyses with the angular power spectrum in
Sec.~\ref{sec:preliminary}. A comprehensive description of our method
follows in Sec.~\ref{sec:method} including the preparation of adequate
Gaussian simulations, the working principle of our spot searching
algorithm, and an error estimation. Our results are presented in
Sec.~\ref{sec:results}. We consider both cut-sky maps (with unreliable
pixels excluded by the KQ75 temperature analysis mask) and the
Internal Linear Combination (ILC) full-sky map, and quantify
deviations from Gaussian simulations. We sum up and conclude in
Sec.~\ref{sec:discussion}.

\section{Preliminary considerations}\label{sec:preliminary}

The most robust comparison between predicted and observed spot
abundances of CMB sky maps relies on simulated maps since analytic
methods can hardly care for complications due to masking and beam
properties. Creating a number of simulated maps and treating them in
exactly the same way as the original map therefore is the clearest
method. Nonetheless, it is instructive to recall some well-known
analytic results that connect the pixel-space analysis to familiar
harmonic space.

The spot abundances in a CMB sky map are dictated by the angular
correlations of temperature fluctuations. The most popular theories
stick to Gaussianity and statistical isotropy.
Then, the ensemble average of the angular correlation between two
directions $(\theta,\varphi)$ and $(\theta',\varphi')$ only depends on
the angle $\Theta$ between them. This leads to the definition of the
angular correlation function
\begin{equation} C(\Theta) = \left< \frac{\Delta T}{\bar
	T}(\theta,\varphi)\times \frac{\Delta T}{\bar T}(\theta',\varphi')
	\right>.
	\label{eq:correlation_fct}
\end{equation}
We can switch to harmonic space by decomposing the temperature field
into spherical harmonics:
\begin{equation}
		\frac{\Delta
		T}{\bar T}(\theta,\varphi)=\sum_{\ell,m} a_{\ell m}
		Y_{\ell m}(\theta,\varphi),
	\label{eq:T_decomposed}
\end{equation}
where the crucial assumption of statistical isotropy implies
\begin{equation}
	\left< a_{\ell m} a_{\ell'm'}^* \right> = \delta_{\ell \ell'}
	\delta_{mm'} C_\ell.
	\label{eq:stat_isotropy}
\end{equation}
So, in this case, all the statistical information is in the
coefficients $C_\ell$, the angular power spectrum. More generally, we
may define
\begin{equation}
	C_\ell = \frac{1}{2 \ell+1} \sum_{m}^{} \left< |a_{\ell m}|^2 \right>.
	\label{eq:Cl_def}
\end{equation}

When searching for spots of a given size, we will average the
temperature fluctuations in regions of that size. These regions are
defined by window functions $W(\theta,\varphi)$. The mean temperature
contrast in such a region is
\begin{equation}
	\Delta T = \int_{}^{} \dderiv\Omega\, \Delta
	T(\theta,\varphi)\, W(\theta,\varphi).
	\label{DTrmsWindow}
\end{equation}
In our sense, a {\it spot} is characterized as follows. When a
threshold $\Delta \mathfrak T$ is fixed, a {\it hot spot} is found if
$\Delta T \ge \Delta \mathfrak T$, whereas a {\it cold spot} is found
if $\Delta T \le - \Delta \mathfrak T$.
The characteristic scale for $\Delta T$ is the mean temperature
contrast for these regions $\Delta T_{rms} = \sqrt{\left< \Delta T^2
\right>}$. Clearly, if $\Delta \mathfrak T \ll \Delta T_{rms}$, most regions
will be spots, if $\Delta \mathfrak T \gg \Delta T_{rms}$, only a few
or none.

\begin{figure}[htb!]
	\begin{center}
		\psfrag{xlabel}[B][c][.8][0]{$\ell$}
		\psfrag{ylabel}[B][c][.8][0]{$W_\ell$}
		\includegraphics[width=.4\textwidth]{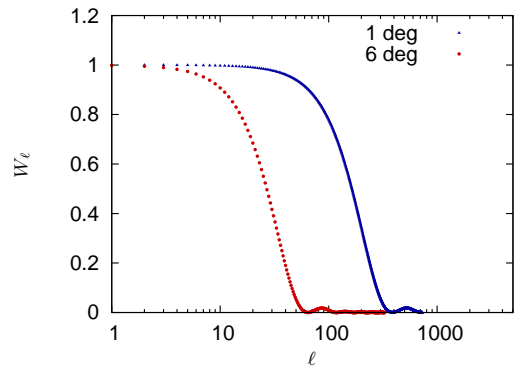}
	\end{center}
	\caption{Coefficients $W_\ell$ for the top hat circle window function
	at scales $a=1^\circ$ (right plot), $6^\circ$ (left plot). The
	plots show which multipoles predominantly determine $\Delta T_{rms}$.
	For smaller angular scale $a$, higher $\ell$ values enter the
	analysis.}
	\label{fig:Wls}
\end{figure}

The transformation to harmonic space can be done by decomposing the
window function $W(\theta,\varphi)$ into spherical harmonics with
coefficients $W_{\ell m}$ and defining
\begin{equation}
	W_\ell = \sum_{m}^{} |W_{\ell m}|^2.
	\label{eq:Wl}
\end{equation}
Together with Eqs.~(\ref{eq:T_decomposed}) and
(\ref{eq:stat_isotropy}), it is straightforward to calculate
\begin{equation}
	\Delta T_{rms}^2 = \sum_{\ell}^{}
	\frac{2 \ell+1}{4\pi}\,C_\ell\,W_\ell\,\bar T^2.
	\label{eq:DTrms}
\end{equation}
This result shows that the mean temperature fluctuation $\Delta
T_{rms}$ is given by the multipoles $C_\ell$ weighted by $W_\ell$. The $W_\ell$
strongly depend on the angular scale of the regions. Their magnitude
will suppress large $\ell$ values corresponding to scales smaller than
the window. By virtue of the addition theorem for spherical
harmonics, we can write
\begin{equation}
	W_\ell = \int \dderiv\Omega\ \int
	\dderiv\Omega'\ 
	W(\theta,\varphi)\, W(\theta',\varphi')\, P_\ell(\cos\Theta).
	\label{Wl_integration}
\end{equation}
This allows us to calculate the $W_\ell$ for a chosen window. An example
is shown in Fig.~\ref{fig:Wls}.

\begin{figure}[htb]
	\begin{center}
		\psfrag{xlabel}[B][c][.8][0]{$a$ [$^\circ$]}
		\psfrag{ylabel}[B][c][.8][0]{$\Delta T_{rms}$ $[\upmu \text K ]$} 
		\includegraphics[width=.4\textwidth]{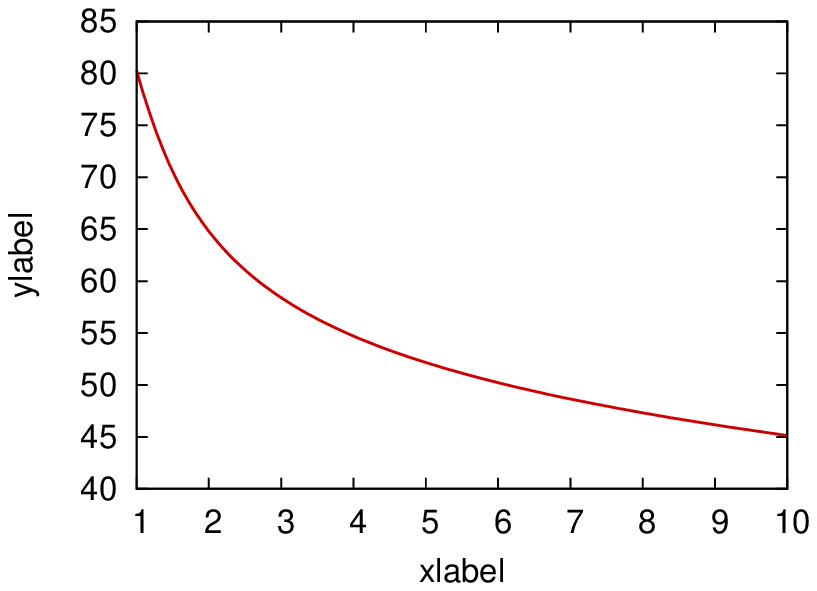}
		\psfrag{xlabel}[B][c][.8][0]{$a$ [$^\circ$]}
		\psfrag{ylabel}[B][c][.8][0]{Relative deviation in \%}
		\includegraphics[width=.4\textwidth]{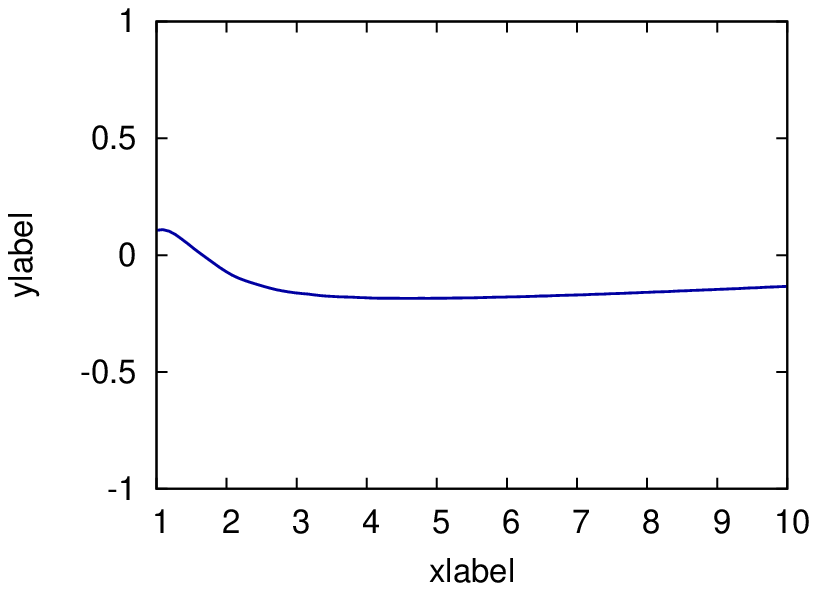}
		\caption{Mean temperature fluctuation for various spot
		sizes and the $\Lambda$CDM power spectrum. The plots for
		circles and squares are visually indistinguishable. The
		difference between the result for circles and the result
		for squares is shown in the second figure.}
		\label{fig:DTrms}
	\end{center}
\end{figure}

In our case, it is adequate to approximate the sphere by the tangent
plane at a region, replacing the direction $(\theta,\varphi)$ by points
$\vec x$ on the plane. For our purposes, it is most convenient to work
with top hat windows because they have clear boundaries. This is the
easiest way to avoid ambiguities arising from overlapping spots.
Exemplary choices may be the top hat circle with window function
\begin{equation}
	W(\vec x) = \frac{1}{\pi R^2}\Theta(R-|\vec x|)
	\label{eq:W_circle}
\end{equation}
or a square with window function
\begin{equation}
	W(\vec x) = \frac{1}{a^2}\, \Theta(a-x_1)\,
	\Theta(a-x_2).
	\label{eq:W_square}
\end{equation}
Following \citet[p. 218]{Durrer08}, we can approximate the $W_\ell$ by
an angular average over the Fourier transform of $W(\vec x)$ which
considerably reduces the computational effort:
\begin{equation}
	W_\ell \approx \frac{1}{2\pi} \int_{0}^{2\pi} \dderiv \alpha \, |\tilde
	W(\vec l)|^2.
	\label{eq:durrer}
\end{equation}
For the aforementioned window functions, we can use this equation to
easily calculate $\Delta T_{rms}$ by Eq.~(\ref{eq:DTrms}). The
results are plotted for the $\Lambda$CDM best-fit power spectrum in
Fig.~\ref{fig:DTrms}. For the sake of comparability, we use the
parameter $a$ which equals the square root of the windows' area; in
the case of squares, it simply is the side length. We also show the
relative deviation due to the different window functions. We conclude
that the result is not sensitive to the exact geometry if the covered
surface area is the same.

\section{Method}\label{sec:method}

Our strategy consists of performing an identical analysis of spot
abundances both for observational maps and maps generated from
simulations of Gaussian fluctuations. For the simulated maps, we use
the best-fit $\Lambda$CDM model and a Gaussian fluctuation model based
on the $C_\ell$ quoted by the WMAP collaboration. The comparison with
maps from observation tests Gaussianity.

Because of the excellent data products of the WMAP team
available at the legacy archive\footnote{http://cmbdata.gsfc.nasa.gov}
and the comprehensive HEALPix
package\footnote{http://healpix.jpl.nasa.gov}~\citep{HEALPIX}, it is
possible to obtain reliable CMB sky maps and to create maps from
Gaussian simulations. We summarize the steps in Sec.~\ref{sec:data}.
We developed an algorithm searching for hot and cold spots (in the
sense of Sec.~\ref{sec:preliminary}) within these temperature sky
maps.  Its working principle and properties are presented in
Sec.~\ref{sec:algorithm}. The treatment of statistical errors is
described in Sec.~\ref{sec:errors}.

\subsection{Maps and data preparation}\label{sec:data}

\begin{figure}[htb]
	\begin{center}
		\subfigure
		{
			\includegraphics[width=.4\textwidth]{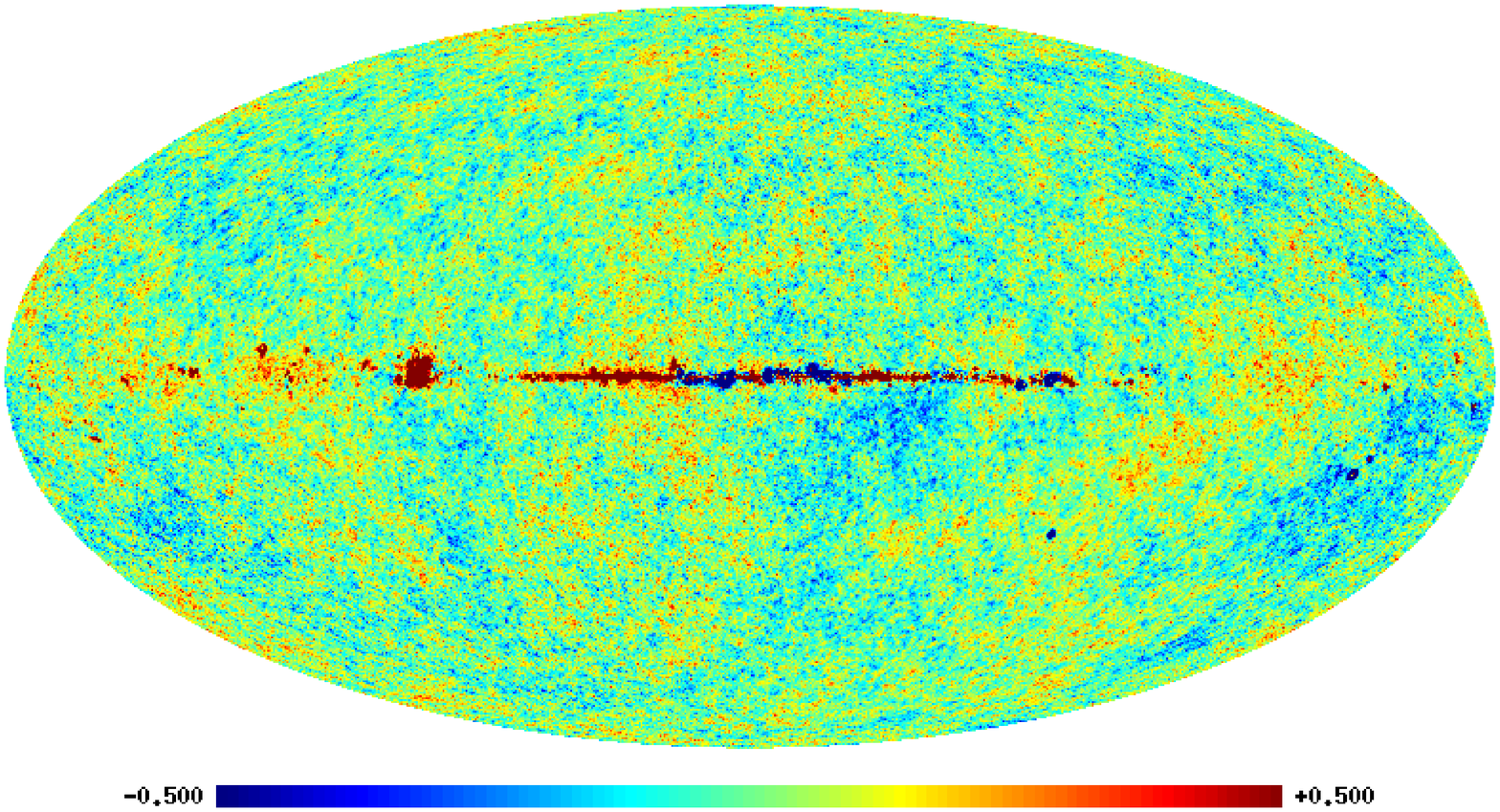}
		}
		\subfigure
		{
			\includegraphics[width=.4\textwidth]{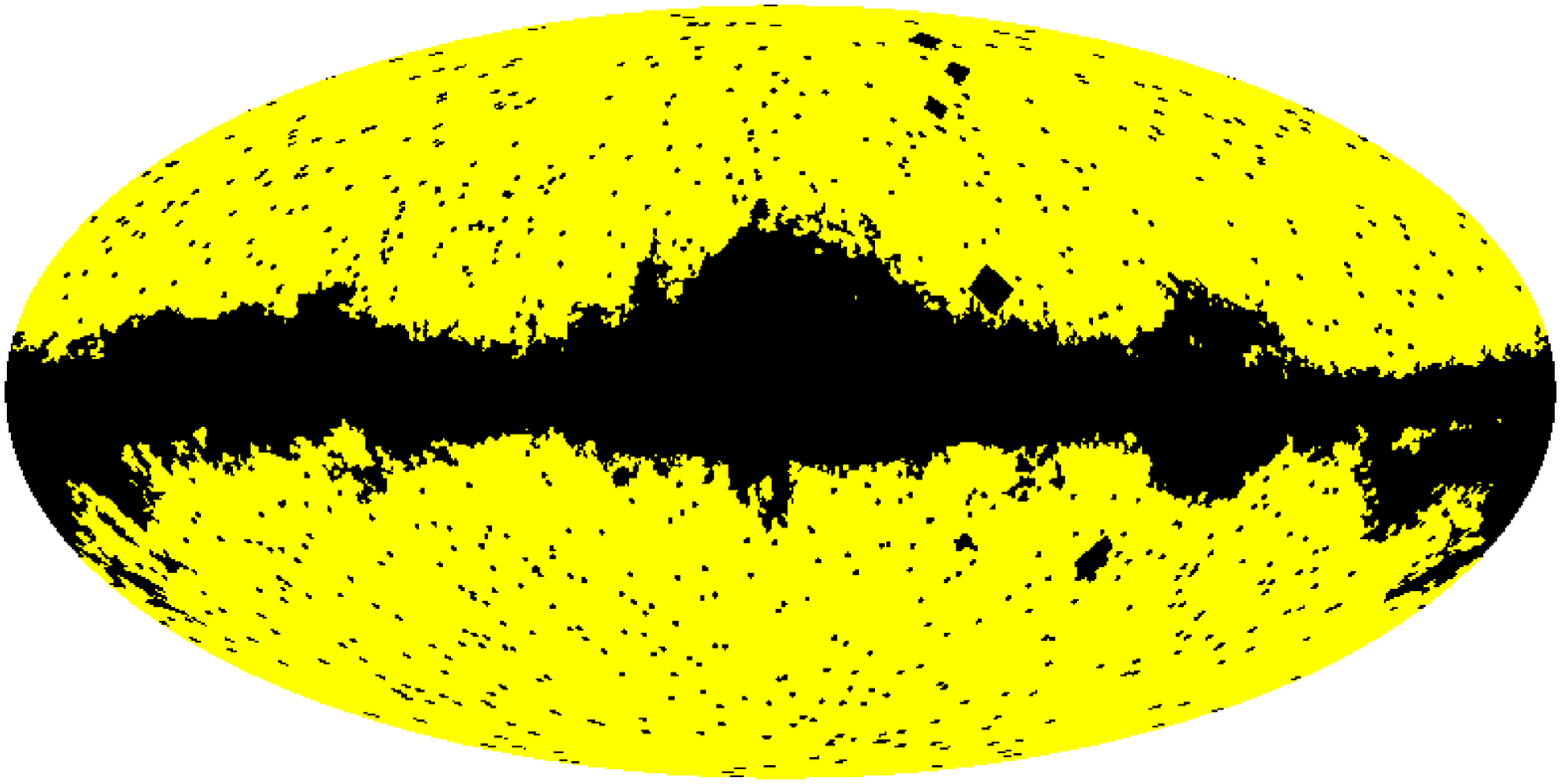}
		}
	\end{center}
	\caption{The foreground-reduced $V$ map (temperature contrast
	in mK) and the KQ75 mask
	cutting out the contaminated galaxy region and point sources.}
	\label{fig:maps}
\end{figure}

Whenever the original signal is to be extracted from CMB data, it is
crucial to minimize the influence of foreground contamination. The
frequency dependence of the foreground components (e.g.,  synchrotron
emission, free-free emission, and thermal dust) allows to reduce the
contamination with the help of various foreground
models~\citep{Gold09}. The WMAP team provides foreground-reduced maps
for the $Q$ (35--46 GHz), $V$ (53--69 GHz), and $W$ (82--106 GHz) bands. Since
the $V$ band has a better signal-to-noise ratio than the $W$ band and is
less foreground contaminated than the $Q$ band~\citep{Hinshaw07}, it is
the natural choice to use the foreground-reduced $V$ map. Further
noise minimization by constructing linear combinations of the maps is
possible but does not affect our analysis which focuses on large
scales. But still, large parts of the temperature map are unreliable
and must be excluded from the analysis. We therefore apply the KQ75
mask, cutting out the contaminated galaxy region and point
sources~\citep{Gold09}. Finally, the residual monopole and dipole are
removed with the HEALPix routine \mbox{\sc remove\_dipole}.
Figure~\ref{fig:maps} shows the foreground-reduced $V$ map and the
KQ75 mask.

Gaussian simulations based on some input $C_\ell$ spectrum and a beam
window function are achieved with the help of the {\sc synfast}
HEALPix facility. These input data can be obtained from the legacy
archive. The power spectra we used are the $\Lambda$CDM best fit and
the original \mbox{WMAP-5} estimate both shown in Fig.~1 of
\citet{Nolta09}.  Subsequently, we will refer to them by
``$\Lambda$CDM'' and ``\mbox{WMAP-5}'' power spectrum for short. We
take care of treating simulated and original maps as equally as
possible.  This necessitates the additional simulation of the
instruments' noise, masking, and removal of monopole and dipole. Since
the WMAP design minimizes noise correlation between neighboring pixels
in a map~\citep{Page03}, it is legitimate to add white noise with the
properties described by the WMAP team at the legacy archive.

When studying possible anisotropy of the CMB, we need a full-sky
(unmasked) map. Since the foreground contaminations usually force us
to mask parts of the sky, it is not a trivial task to reconstruct the
full-sky CMB signal. However, the WMAP team tries to tackle this job
by combining the measurements of all bands and merge them into a
single (ILC) map of the full sky~\citep{Gold09}. The applied procedure
is independent of foreground models but has the disadvantage of being
doubtful on scales below approximately $10^\circ$ according to the
WMAP product description at the legacy archive. But since we are
lacking any better alternative, we employ the 5-year WMAP ILC map for
full-sky analyses.

\subsection{Spot searching algorithm}\label{sec:algorithm}

The primary goal of the algorithm is to count hot and cold spots in
CMB sky maps on various scales and temperature contrasts. A typical
application will be to plot spot abundances against the threshold on
the temperature contrast $\Delta \mathfrak T$ for a specific angular
scale. This application directly imposes several features the
algorithm should have:
\begin{packed_itemize}
	\item[(i)] It must define {\it sectors on the sphere of equal
		surface area} (for some desired scale). Their mean temperature
		contrasts will decide whether they are counted as spots.
	\item[(ii)] The areas must be chosen such that one can {\it
		smoothly scan} through the map. Between two distinct areas,
		there must exist many others allowing for a smooth transition.
	\item[(iii)] Double counting of spots has to be excluded. The
		easiest way to achieve this is working with {\it top hat
		windows} which have clear boundaries. Overlapping spots will be
		counted as a single.
	\item[(iv)] For a statistically satisfactory comparison between
		observed and simulated CMB maps, the algorithm will have to
		analyze many sky maps. Given the huge amount of data, one has
		to implement the algorithm carefully in order to make this
		{\it numerically tractable}.
\end{packed_itemize}
The algorithm is designed such that it allows for an approximate
pixelization of the sphere into distinct areas of a given scale.
Calculating their temperature contrasts determines the mean
temperature fluctuation $\Delta T_{rms}$ on that scale. By virtue of
the ergodic theorem, this is a good estimate for the ensemble average
introduced in Sec.~\ref{sec:preliminary}.

\subsubsection{Working principle}\label{sec:principle}

The first task is to define the sectors $S$ of equal surface area on
the sphere satisfying the requirements explained above. We choose them
to be intersections of latitude and longitude rings. A latitude ring
$\mathcal R^{lat}$ consists of all points between two latitude angles
$\theta_0$ and $\theta_1$, a longitude ring $\mathcal R^{lon}$ of all
points between two longitude angles $\varphi_0$ and $\varphi_1$. The
rings have two nice properties. First, as needed for spot searching,
one can smoothly go from one ring to any other ring by smoothly
changing its boundary angles; second, as needed for calculating
$\Delta T_{rms}$, it is an easy task to discretize a sphere into
distinct rings. Since sectors are intersections $S = \mathcal R^{lat}
\cap \mathcal R^{lon}$, they share these properties. We thereby
satisfy the requirement of smooth scanning to all directions.

We impose [meeting the requirement~(i) above] equal area $A$ for all
sectors:
\begin{equation}
	A = \int_{\mathcal S}^{} \dderiv \Omega =
	\int_{\varphi_0}^{\varphi_1}\dderiv\varphi\
	\int_{\theta_0}^{\theta_1} \dderiv\theta\ \sin\theta.
	\label{eq:equal_area}
\end{equation}
Once we
have decided to define sectors like this, we still have some freedom
to choose the boundaries $\theta_0$, $\theta_1$, $\varphi_0$,
$\varphi_1$. In order to avoid the influence of small scales, we must
reasonably choose the sectors such that they are by no means
degenerated. We therefore fix this freedom by adding another
constraint. For any sector $\mathcal S$, the boundary lines in the
north-south direction and the longer boundary in the east-west
direction are chosen to be of equal length:
\begin{equation}
	(\varphi_1-\varphi_0)\, \sin\theta_* =
	(\theta_1-\theta_0).
	\label{eq:equal_length}
\end{equation}
On the northern hemisphere $\theta_*=\theta_1$, on the southern
hemisphere $\theta_*=\theta_0$.  Note that these sectors behave well.
In the limiting case near the equator, they correspond to squares in
flat space. At the poles ($\theta_0=0$ or $\theta_1=\pi$), they become
equilateral triangles.

In practice though, the temperature field is not given as a smooth
function of $\theta$ and $\varphi$. The WMAP temperature sky maps are
lists assigning a temperature contrast $\Delta T_i$ to each HEALPix
pixel $p_i$. The mapping $p_i \mapsto (\theta,\varphi)$ is given in
the form of a table. But since our approach defines sectors by means
of angles, we need the reverse. Given the list $p_i \mapsto
(\theta,\varphi)$, finding the appropriate pixel $p_i$ for given
angles $(\theta,\varphi)$ corresponds to searching through the list.
Whereas searching in an unsorted list is very expensive, an adequate
sorting may considerably reduce the effort. The algorithm performs the
following steps starting at the north pole \mbox{$\theta_0=0$}:
\begin{packed_enum}
	\item For given $\theta_0$ and area $A$, calculate $\theta_1$ and
		$\Delta \varphi$ by solving Eqs.~(\ref{eq:equal_area})
		and (\ref{eq:equal_length}).
	\item Collect the pixels $\{p_i\}$ belonging to the latitude ring
		$\mathcal R^{lat}$ between $\theta_0$ and $\theta_1$.
		This can be done efficiently if the map was prepared by
		transforming to sorted latitude angles (HEALPix {\sc ring}
		ordering).
	\item Using a fast routine, sort the list $\{p_i\}$ with respect
		to longitude angles. This new sorting allows one to directly
		identify the pixels out of $\{p_i\}$ belonging to a longitude
		ring $\mathcal R^{lon}$ with boundaries $\varphi_0$ and
		$\varphi_1$---these pixels form the sector $\mathcal S =
		\mathcal R^{lat} \cap \mathcal R^{lon}$. Start at
		$\varphi_0=0$ and $\varphi_1=\Delta\varphi$ and smoothly scan
		(by increasing $\varphi_0$, $\varphi_1$
		by a small step size $h$)
		through all longitude rings. For every sector, calculate the
		sector's mean temperature contrast $\Delta T$ by averaging
		over the pixel values $\Delta T_i$ and compare it with
		the threshold $\Delta \mathfrak T$. If it exceeds the
		threshold, count a spot if the sector does not overlap with a
		previously found spot.
	\item Choose the next ring by slightly increasing $\theta_0 
		\mapsto \theta_0 + h$. It
		is profitable to exploit the fact that the sorting for
		longitude angles (point 3) need not be repeated completely. The
		algorithm saves the previous sorting and uses it for a
		presort such that as much information is transferred as
		possible.
\end{packed_enum}
Having increased the threshold $\Delta\mathfrak T$, again searching
for spot abundances in a map can be optimized by noticing that spots
at a higher threshold cannot be found where there were not spots at a
lower threshold. Our algorithm can focus on areas around previously
found spots once this becomes advantageous.

\begin{figure}[htb]
	\begin{center}
		\includegraphics[width=.4\textwidth]{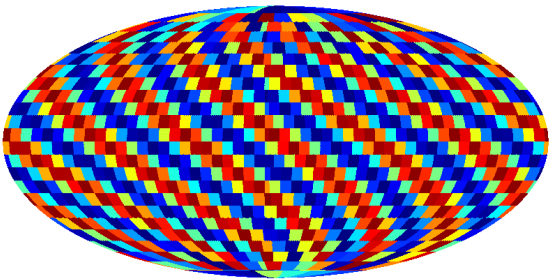}
	\end{center}
	\caption{Exemplary decomposition of the sky into $N_{sec}$
	distinct sectors $\mathcal S_j$ for measuring $\Delta T_{rms}$.
	For searching spots, the algorithm analyzes many more sectors
	$\mathcal S$ (those in between, sharing pixels with the
	illustrated sectors $\mathcal S_j$). Nonetheless, $N_{sec}$ limits
	the maximum number of spots since overlapping spots are not
	multiply counted.}
	\label{fig:sectors}
\end{figure}

If we slightly adapt the algorithm, we can use it to measure $\Delta
T_{rms}$. Now, the algorithm jumps between distinct sectors instead of
smoothly transforming them. The distinct sectors are visualized in
Fig.~\ref{fig:sectors}.
In every distinct sector, the mean temperature fluctuation is
calculated. The squares are averaged to give $\Delta T_{rms}$.
Although the shapes of the sectors vary, the results of
Sec.~\ref{sec:preliminary} ensure that $\Delta T_{rms}$ is only
marginally affected.

\subsubsection{Treatment of masked maps}\label{sec:mask_treatment}

The sectors defined by our algorithm may include none, some, or many
masked pixels. We must define selection rules determining which
sectors are to be included in the statistics. We used the following
two rules. The most restrictive choice is to only consider sectors
with no mask overlap ({\it strict selection} for short). These sectors
will only contain reliable pixels. But since especially on large
scales, only a minority of sectors will belong to this group, bad
statistics are the price to pay. The alternative choice is to also
consider sectors with a slight mask overlap ({\it tolerant
selection}). This is a compromise between good statistics on the one
hand and reliable results on the other. We typically allow for $5\%$
masked area within a sector which guarantees that usually the majority
of sectors fall into this group. In any case, we emphasize that masked
pixels, even if included in the statistics, are assigned zero
temperature fluctuation. This will avoid misinterpreting foregrounds
as a CMB signal. Note however, that the pixels of zero temperature
fluctuation reduce $\Delta T_{rms}$. For comparisons between observed
maps and Gaussian simulations, we employ the tolerant selection for
the sake of better statistics; the comparison is still trustworthy.

\subsubsection{Alternative shapes}\label{sec:shapes}

\begin{figure}[htb]
	\begin{center}
		\psfrag{xlabel}[B][c][.8][0]{Threshold
		$\Delta\mathfrak T$ [$\upmu \text{K}$]}
		\psfrag{ylabel}[B][c][.8][0]{Abundances of spots} 
		\includegraphics[width=.45\textwidth]{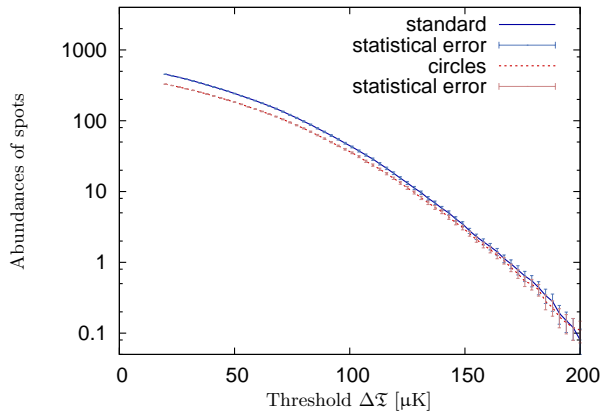}
	\end{center}
	\caption{Mean spot abundances in 100 simulated $\Lambda$CDM
	full-sky maps showing the results for different window
	functions of scale $a=\sqrt{A}=6^\circ$.}
	\label{fig:circles}
\end{figure}

The algorithm works with the shapes defined in
Sec.~\ref{sec:principle} and illustrated in Fig.~\ref{fig:sectors}.
But we can easily treat other shapes by embedding them into the
previous sectors. This corresponds to a multiplication of the previous
window function $W_0$ with the window function $W_1$ of the desired
shape where $W_0$ must be large enough to ensure $W_0\equiv 1$ where
$W_1$ is non-zero. The condition (\ref{eq:equal_area}) of equal area
now concerns the new shape and reads
\begin{equation}
	\int_{\varphi_0}^{\varphi_1}\dderiv\varphi\
	\int_{\theta_0}^{\theta_1} \dderiv\theta\ \sin\theta \ W_1(\theta,\varphi)
	= A.
	\label{eq:equal_area_shapes}
\end{equation}
As an example, we compare the standard shape with top hat circles
of equal area [cf. Eq.~(\ref{eq:W_circle})] and plot the result in
Fig.~\ref{fig:circles}.
For low thresholds, the abundances are systematically higher for the
standard window function. This is due to the fact that circles do not
exhaust the area without space in between. The effect becomes
important where many spots are found and overlapping is frequent but
disappears for large thresholds where the results agree.

\subsubsection{Step size dependence}\label{sec:stepsize}

\begin{figure}[htb]
	\begin{center}
		\psfrag{xlabel}[B][c][.8][0]{Step size $h$ [$^\circ$]}
		\psfrag{ylabel}[B][c][.8][0]{Abundances of spots} 
		\includegraphics[width=.4\textwidth]{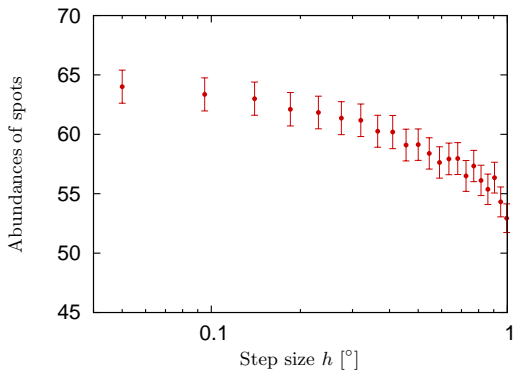}
	\end{center}
	\caption{Mean spot abundances for a fixed threshold (80 $\upmu
	\text{K}$) against a varying step size. 100 masked $\Lambda$CDM
	simulated maps were scanned for $a=6^\circ$,
	the error bars quantify the statistical error.}
	\label{fig:stepsize}
\end{figure}

In the ideal case, the boundary angles of the sectors would vary in a
perfectly smooth manner when searching for spots in a map. But
numerically, we have to choose a finite step size $h$ (introduced in
Sec.~\ref{sec:principle}). A good choice
of $h$ balances sensitivity and numerical effort.
Figure~\ref{fig:stepsize} shows detected spot abundances against $h$
in simulated maps. We chose $h=0.3^\circ$ for which we conclude that
our sensitivity to detect spots is satisfactory.

\subsection{Errors and cosmic variance}\label{sec:errors}

There are statistical uncertainties simply due to the finite number of
Gaussian simulations. Moreover, the CMB signal itself can be regarded
as the outcome of a statistical process. It is therefore subject to
statistical variation, quantified by the concept of cosmic variance.

Let us assume that $N$ Gaussian maps are analyzed for spots (area
and threshold fixed). If $n^{(k)}$ spots are detected in map $k$, the
mean spot abundance is
\begin{equation}
	\bar n = \sum_{k=1}^{N}\frac{n^{(k)}}{N}.
	\label{eq:mean_abundance}
\end{equation}
The statistical uncertainty of the mean value $\bar n$ and the
statistical deviation of the single values $\bar n^{(k)}$ are
\begin{equation}
	\sigma_{\bar n}^2 = \frac{\sum_{k=1}^{N}(n^{(k)}-\bar n)^2}{N
	(N-1)}, \ \sigma_{n^{(k)}}^2 = N \, \sigma_{\bar n}^2.
	\label{eq:errors_abundance}
\end{equation}
The same procedure applies if we measure the mean temperature
fluctuations $\Delta T_{rms}^{(k)}$ in the maps and calculate a mean
value $\Delta \bar T_{rms}^{(k)}$.

We now consider cosmic variance. When we observe a spot abundance $n$,
we must expect a certain deviation from the theoretically predicted
ensemble average $\left< n \right>$. The expectation value of this
deviation, $\sigma_n^2 = \left< (n - \left< n \right>)^2 \right>$,
quantifies cosmic variance. For a very large number $N$ of simulated
maps, we may replace the ensemble average by an averaging over the set
of simulations. We then obtain $\sigma_n \approx \sigma_{n^{(k)}}$
with the latter calculated according to
Eq.~(\ref{eq:errors_abundance}). This can be done equally for the mean
temperature contrast $\Delta T_{rms}$. Whenever we specify cosmic
variance (e.g., in the form of error bars), we estimated it by this
method.

\section{Results}\label{sec:results}

The application of the spot-searching algorithm described in
Sec.~\ref{sec:method} shows that the standard model $\Lambda$CDM
together with Gaussianity predicts more large-scale hot and cold spots
than are actually present in cut-sky \mbox{WMAP-5} data (see
Sec.~\ref{sec:cut-sky}). Removing the quadrupole or using the original
\mbox{WMAP-5} $C_\ell$ (instead of the $\Lambda$CDM fit) considerably reduces the
discrepancies. While only $0.16\%$--$0.62\%$ of Gaussian $\Lambda$CDM
simulations fall below the observed mean temperature fluctuations on angular
scales of $4^\circ$--$8^\circ$, this increases to $2.5\%$--$8\%$ if the quadrupole is removed. 
We also investigate full-sky maps in
Sec.~\ref{sec:ilc} and modifications of the first multipoles in
Sec.~\ref{sec:modified}.

\subsection{Cut-sky maps}\label{sec:cut-sky}

\begin{figure}[htb]
	\begin{center}
		\psfrag{xlabel}[B][c][.8][0]{Threshold
		$\Delta\mathfrak T$ [$\upmu \text{K}$]}
		\psfrag{ylabel}[B][c][.8][0]{Abundances of hot spots} 
		\includegraphics[width=.45\textwidth]{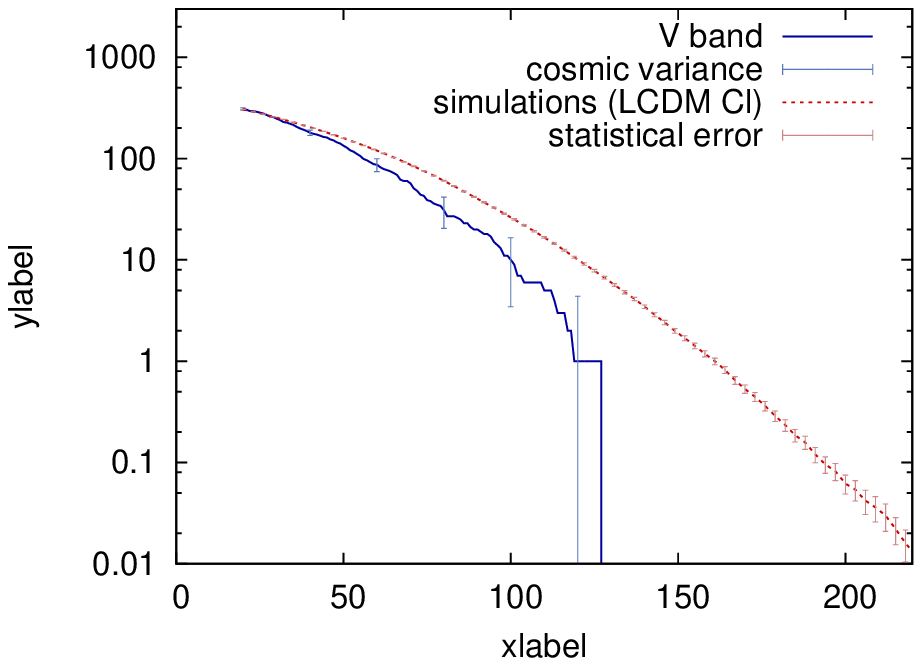}
		\psfrag{xlabel}[B][c][.8][0]{Threshold
		$\Delta\mathfrak T$ [$\upmu \text{K}$]}
		\psfrag{ylabel}[B][c][.8][0]{Abundances of cold spots}
		\includegraphics[width=.45\textwidth]{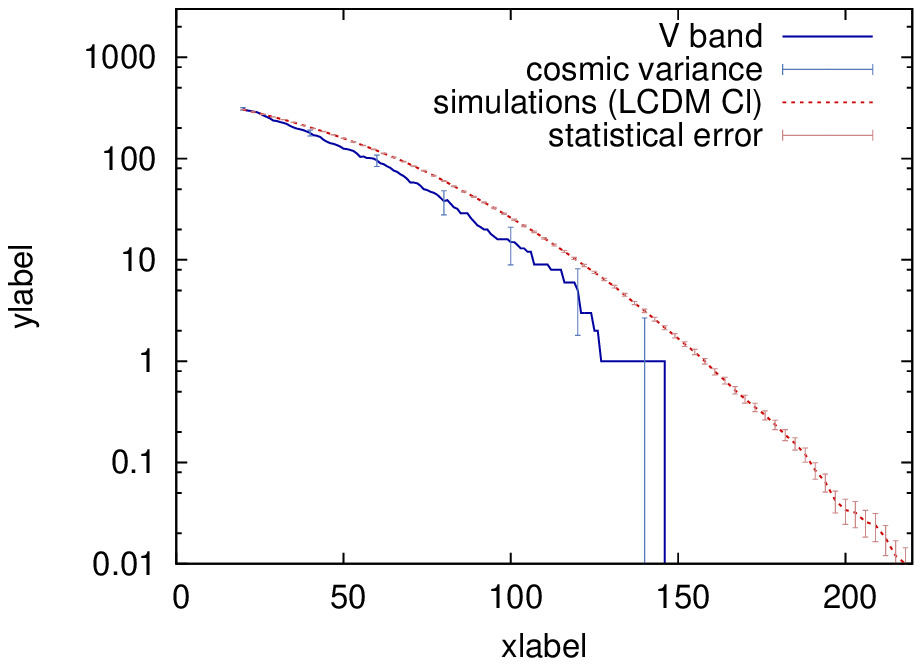}
		\caption{Spot abundances in the CMB sky (with cosmic variance) as compared to 500
		$\Lambda$CDM simulations (with statistical errors) on an angular scale of
		$a=6^\circ$. The fractions of Gaussian simulations with
		smaller values of $s$ [Eq.~(\ref{eq:s})] are $p_s^\text{hot} =
		0.2\%$ and $p_s^\text{cold} = 1.8\%$.}
		\label{fig:spots_V_LCDM}
	\end{center}
\end{figure}

\begin{figure}[htb]
	\begin{center}
		\psfrag{xlabel}[B][c][.8][0]{Threshold
		$\Delta\mathfrak T$ [$\upmu \text{K}$]}
		\psfrag{ylabel}[B][c][.8][0]{Abundances of hot spots}
		\includegraphics[width=.45\textwidth]{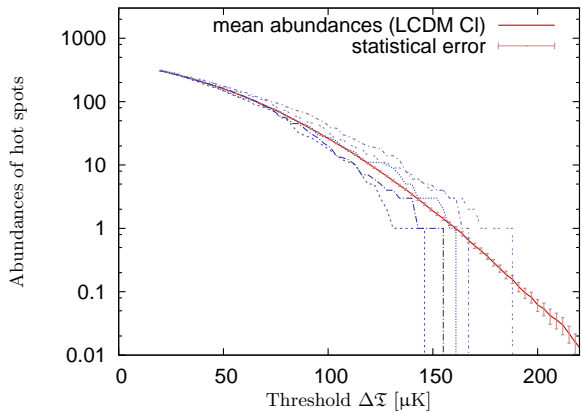}
		\caption{Spot abundances in five randomly chosen Gaussian simulations based on the $\Lambda$CDM best-fit power spectrum and the mean curve from Fig.~\ref{fig:spots_V_LCDM} (hot spots). }
		\label{fig:gaussians}
	\end{center}
\end{figure}

The spots' size is characterized by their area $A$. We use the
parameter $a=\sqrt{A}$ to specify the angular scale of this size.
Since on the one hand, we aim at large scales, and on the other hand,
we want reasonable statistics, we are forced to find a compromise. We
chose an angular scale $a=6^\circ$. The spot abundances of the
\mbox{WMAP-5} $V$ map and 500 Gaussian $\Lambda$CDM simulations
(created as described in Sec.~\ref{sec:data}) are found for varying
threshold $\Delta \mathfrak T$. The HEALPix resolution parameter of
the maps is 8, corresponding to $N_{pix}=12\times256^2=786,432$
pixels.  Statistical uncertainties and cosmic variance are displayed
as error bars even though the spot abundances for different thresholds
are of course correlated. The results for hot and cold spots are
plotted in Fig.~\ref{fig:spots_V_LCDM}.
The striking feature of the plots is the discrepancy
between theory and observation. They only agree in the limit of very
small thresholds $\Delta\mathfrak T$ where it is obvious that 
almost every area is counted as a spot anyway. The discrepancy is seemingly
more drastic for hot spots. In the plot for cold spots, it is seen
that there is one considerable cold spot nearly reaching $150\,\upmu \text{K}$.
But even this spot does not surpass the $\Lambda$CDM prediction. We
note that this spot is localized in the region of the famous Vielva
cold spot~\citep{Vielva03}. It is insightful to look at the spot distributions of single Gaussian simulations 
in order to get an impression of their typical behavior. 
Five examples are plotted in Fig.~\ref{fig:gaussians}.  

Because of the strong correlation between the spot abundances $n_i$ for
different thresholds $\Delta\mathfrak T_i$, it is difficult to judge
the significance of the discrepancies by eye. A possible quantity that
can be used for a comparison of the observed CMB map with Gaussian
simulations is obtained by summing up the spot abundances at different
thresholds,
\begin{equation}
	s = \sum_{i}^{} n_i,
	\label{eq:s}
\end{equation}
where the lowest threshold included is chosen to be the characteristic
scale $\Delta\bar T_{rms}$. We denote the fraction of Gaussian
simulations $k$ with $s^{(k)}$ smaller than found in the $V$ map by
$p_s$. For the spot abundances shown in Fig.~\ref{fig:spots_V_LCDM},
we find $p_s^\text{hot} = 0.2\%$ for hot spots and $p_s^\text{cold} =
1.8\%$ for cold spots.

\begin{figure}[htb!]
	\begin{center}
		\psfrag{xlabel}[B][c][.8][0]{Angular scale $a$
		[$^\circ$]}
		\psfrag{ylabel}[B][c][.8][0]{$\Delta T_{rms}$ [$\upmu
		\text{K}$]} 
		\includegraphics[width=.45\textwidth]{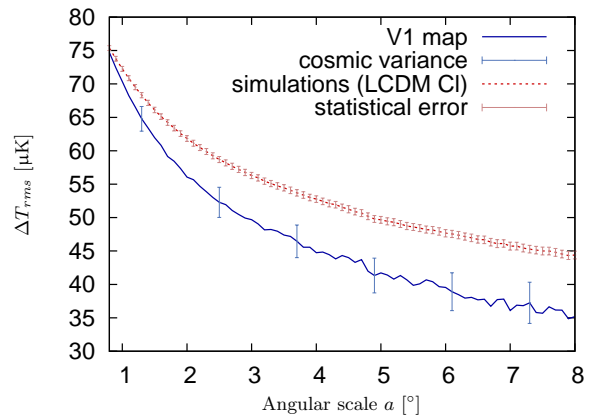}
		\caption{The mean temperature fluctuation for
		different angular scales $a$ in 50 Gaussian $\Lambda$CDM
		simulations and the $V$1 map.}
		\label{fig:sigmas_original}
	\end{center}
\end{figure}

The discrepancies are reflected in the mean temperature fluctuation
$\Delta T_{rms}$ which on large scales is higher in $\Lambda$CDM
simulations than in the observed CMB sky. We have simulated 5000
$\Lambda$CDM maps and compared their mean temperature fluctuations to
the value of the $V$ map. We employed the tolerant selection of our
algorithm (see Sec.~\ref{sec:mask_treatment}).  For $a=6^\circ$, we
find the mean value $\Delta T_{rms} = 39.4\,\upmu \text{K}$ for the $V$ map,
as compared to the mean value $\Delta \bar T_{rms} = 47.9 \pm 0.1 \,
\upmu \text{K}$ for $\Lambda$CDM, where the error is only statistical while
cosmic variance amounts to $4.2\, \upmu \text{K}$. Only a fraction $p=0.6\%$ of
the simulations had a smaller $\Delta T_{rms}$ than the $V$ map. This
does not improve at other large angular scales which can be seen in
Table~\ref{tab:significance}.
\begin{table}
	\begin{center}
	\caption{The fraction $p$ of Gaussian $\Lambda$CDM simulations
	with a $\Delta T_{rms}$ smaller than found in the $V$ map on the
	angular scale $a$.}
	\begin{tabular}{ccccc}
		\hline
		Scale $a$ & & & & Fraction $p$ \\ \hline
		$4^\circ$ & & & & $0.50\%$ \\
		$5^\circ$ & & & & $0.62\%$ \\
		$6^\circ$ & & & & $0.60\%$ \\
		$7^\circ$ & & & & $0.16\%$ \\
		$8^\circ$ & & & & $0.36\%$ \\ \hline
	\end{tabular}
	\label{tab:significance}
\end{center}
\end{table}
It is interesting to see how this behavior changes when going to
smaller scales. However, the results on smaller scales (approaching
$1^\circ$) become sensitive to noise and beam properties. Since the
WMAP team offers the latter for the differencing assemblies $V$1 and
$V$2~\citep{Hill09} instead of the combined $V$ map, it is the easiest to
switch to the $V$1-band map and simulations thereof. The impact on large
scales is negligible. Figure~\ref{fig:sigmas_original} shows $\Delta
T_{rms}$ against the scale $a$ for the $V$1 map and $\Lambda$CDM
simulations (again with tolerant selection).
We see that the deviations decrease when going to smaller scales. This
is also suggested by the $C_\ell$ spectrum which is in good agreement
with the $\Lambda$CDM fit for large $\ell$ which dominate on small
scales. But still, \citet{Monteserin07} find a too low {\it CMB variance} 
which approximately corresponds to $\Delta T_{rms}$ on scales even smaller
than $1^\circ$.

For the results in Fig.~\ref{fig:sigmas_original}, we used the highest
available HEALPix resolution 9 corresponding to $N_{pix}=12\times 512^2 = 3,145,728$
pixels in a map. As stated above, the plots are highly influenced by the beam function
and noise. The beam function acts as an extra window function which
suppresses the growth of $\Delta T_{rms}$ for small scales. The white
noise instead leads to a diverging $1/a$ behavior on the smallest
scales (with an effective pixel noise amplitude $\sigma_{pix}$ and the
number of pixels $N_a = N_{pix}\times a^2/4\pi$ within a sector of
scale $a$, the noise contribution will be
$\Delta T_{rms}^\text{noise}=\sigma_{pix}/\sqrt{N_a}\propto 1/a$).

On large scales, the first multipoles of the $C_\ell$ spectrum play an
important role (see, e.g., Fig.~\ref{fig:Wls}). It is therefore a
natural idea to suspect the well-known quadrupole
anomaly~\citep{Hinshaw07} to be responsible for the observed
discrepancies. We check this by repeating the analysis after removing
the quadrupole from the $\Lambda$CDM simulations as well as the
observed CMB map. The results, summarized in
Table~\ref{tab:remquadrupole}, confirm the influence of the quadrupole
anomaly. Now, the fractions $p$ of Gaussian $\Lambda$CDM simulations
reach $p=7.3\%$ for $a=6^\circ$. These numbers still do not show good
agreement, but they are not statistically significant anymore.
\begin{table}
	\begin{center}
	\caption{The fraction $p$ of $1000$ Gaussian $\Lambda$CDM simulations
	with a $\Delta T_{rms}$ smaller than found in the $V$ map on the
	angular scale $a$, after removing the quadrupole from the maps.}
	\begin{tabular}{ccccc}
		\hline
		Scale $a$ & & & & Fraction $p$ \\ \hline
		$4^\circ$ & & & & $6.5\%$ \\
		$5^\circ$ & & & & $8.0\%$ \\
		$6^\circ$ & & & & $7.3\%$ \\
		$7^\circ$ & & & & $2.5\%$ \\
		$8^\circ$ & & & & $6.4\%$ \\ \hline
	\end{tabular}
	\label{tab:remquadrupole}
\end{center}
\end{table}

\begin{figure}[htb]
	\begin{center}
		\psfrag{xlabel}[B][c][.8][0]{Threshold
		$\Delta\mathfrak T$ [$\upmu \text{K}$]}
		\psfrag{ylabel}[B][c][.8][0]{Abundances of hot spots} 
		\includegraphics[width=.45\textwidth]{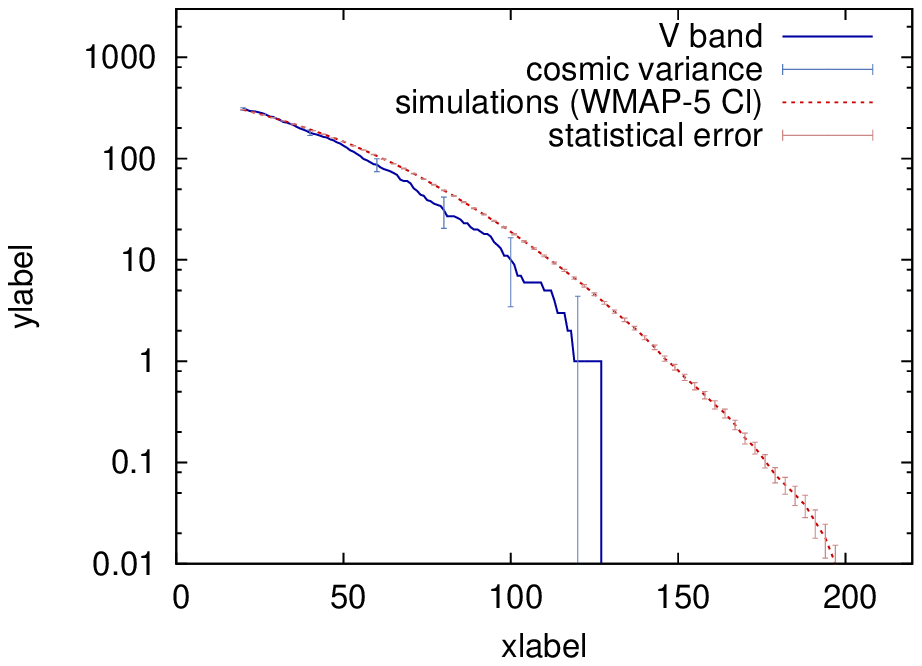}
		\psfrag{xlabel}[B][c][.8][0]{Threshold
		$\Delta\mathfrak T$ [$\upmu \text{K}$]}
		\psfrag{ylabel}[B][c][.8][0]{Abundances of cold spots}
		\includegraphics[width=.45\textwidth]{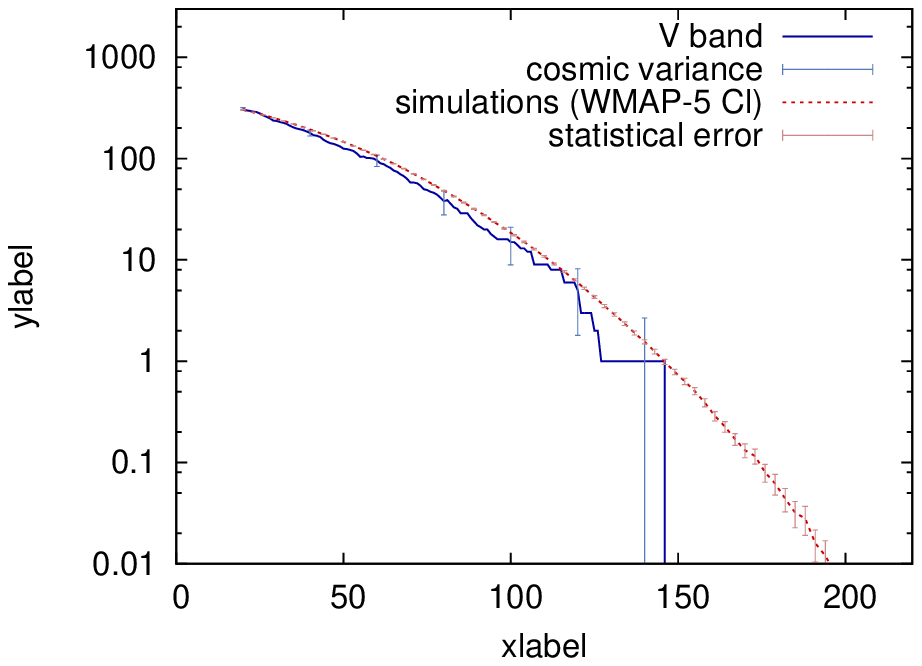}
		\caption{Spot abundances in the CMB sky (with cosmic variance)
		as compared to 500 simulations (with statistical errors) based
		on the \mbox{WMAP-5} $C_\ell$ spectrum on an angular scale of
		$a=6^\circ$. The fractions of Gaussian simulations with
		smaller values of $s$ [Eq.~(\ref{eq:s})] are $p_s^\text{hot} =
		3.4\%$ and $p_s^\text{cold} = 13.2\%$.}
		\label{fig:spots_V_measured}
	\end{center}
\end{figure}

We now investigate whether there are still discrepancies if we compare
the observed $V$ map with Gaussian simulations based on the
original \mbox{WMAP-5} $C_\ell$ spectrum rather than the $\Lambda$CDM best fit.
This tests whether the observed map is a typical Gaussian realization
of the \mbox{WMAP-5} power spectrum. Figure~\ref{fig:spots_V_measured} shows
the spot abundances.
The effect arising from changing the power spectrum is clearly visible
and reduces the discrepancies to some extent. But although closer to
the spot abundances in the observed cut-sky CMB map, the numbers of
hot and cold spots are still too high. Again, this is
reflected in the fact that most simulated maps have a larger $\Delta
T_{rms}$ than the $V$ map. Even though the values, listed in
Table~\ref{tab:significance_measured}, are less drastic, we emphasize
that the \mbox{WMAP-5} estimation of the $C_\ell$ relies on similar data, i.e.,
cut-sky CMB maps. If the observed CMB map was a typical Gaussian
realization of the extracted $C_\ell$ spectrum, we would expect
agreement.
\begin{table}
	\begin{center}
	\caption{The fraction $p$ of 1000 Gaussian simulations
	(\mbox{WMAP-5} $C_\ell$)
	with a $\Delta T_{rms}$ smaller than found in the $V$ map on the
	angular scale $a$.}
	\begin{tabular}{ccccc}
		\hline
		Scale $a$ & & & & Fraction $p$ \\ \hline
		$4^\circ$ & & & & $4.2\%$ \\
		$5^\circ$ & & & & $5.4\%$ \\
		$6^\circ$ & & & & $5.8\%$ \\
		$7^\circ$ & & & & $2.4\%$ \\
		$8^\circ$ & & & & $4.1\%$ \\ \hline
	\end{tabular}
	\label{tab:significance_measured}
\end{center}
\end{table}

Bearing in mind, however, that power spectra refer to the full sky
whereas we only look at regions outside the mask, an explanation could
be that the missing spots were located in the hidden part of the sky.
In the next section, we investigate whether the \mbox{WMAP-5} ILC map
indicates this violation of isotropy.

\subsection{ILC full-sky map}\label{sec:ilc}

\begin{figure}[htb!]
	\begin{center}
		\subfigure{
		\psfrag{xlabel}[B][c][.8][0]{Threshold
		$\Delta\mathfrak T$ [$\upmu \text{K}$]}
		\psfrag{ylabel}[B][c][.8][0]{Abundances of cold spots} 
		\includegraphics[width=.44\textwidth]{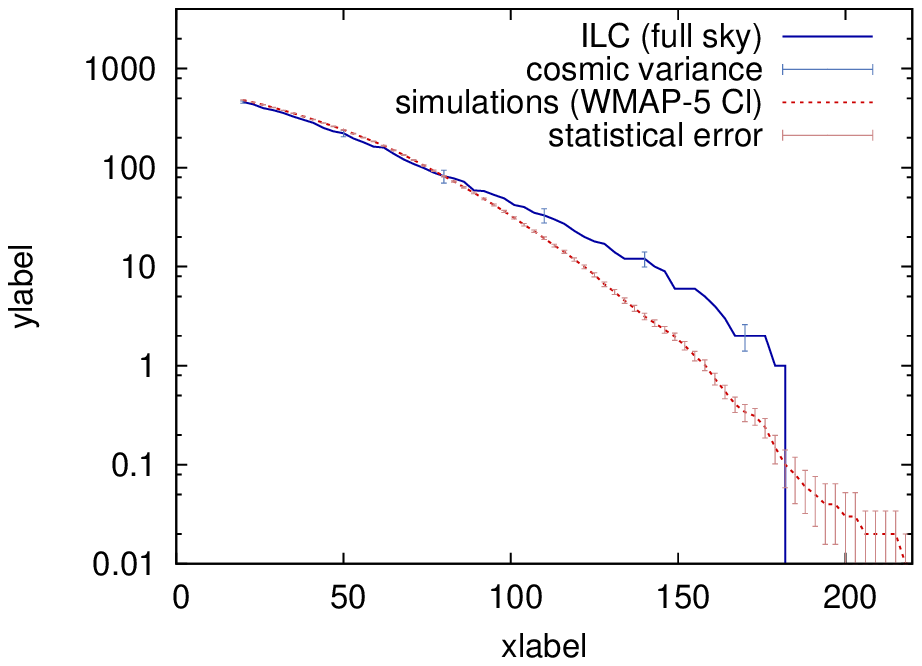}
		}
		\subfigure{
		\psfrag{xlabel}[B][c][.8][0]{Threshold
		$\Delta\mathfrak T$ [$\upmu \text{K}$]}
		\psfrag{ylabel}[B][c][.8][0]{Abundances of cold spots}
		\includegraphics[width=.44\textwidth]{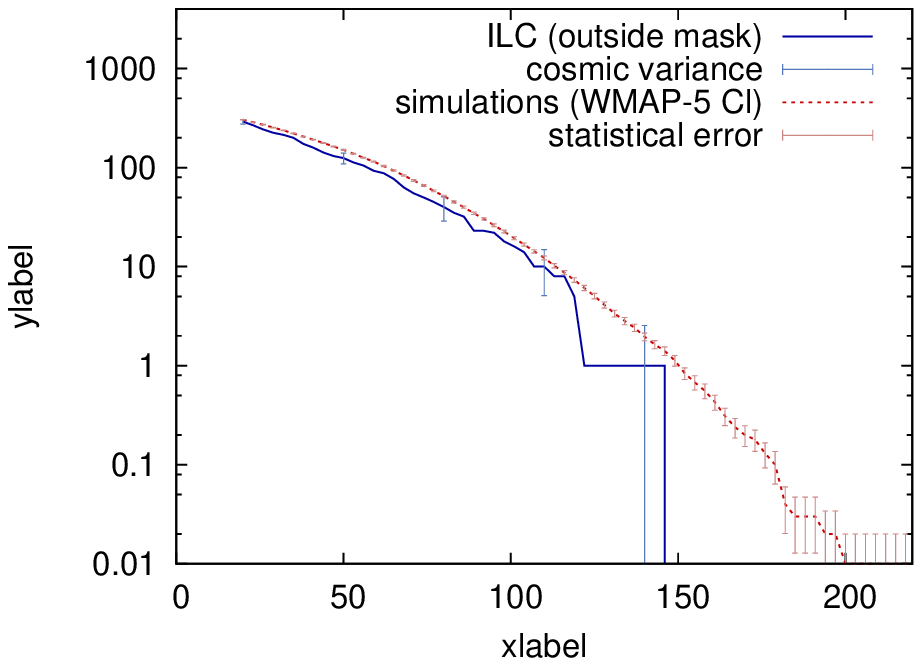}
		}
		\subfigure{
		\psfrag{xlabel}[B][c][.8][0]{Threshold
		$\Delta\mathfrak T$ [$\upmu \text{K}$]}
		\psfrag{ylabel}[B][c][.8][0]{Abundances of cold spots}
		\includegraphics[width=.44\textwidth]{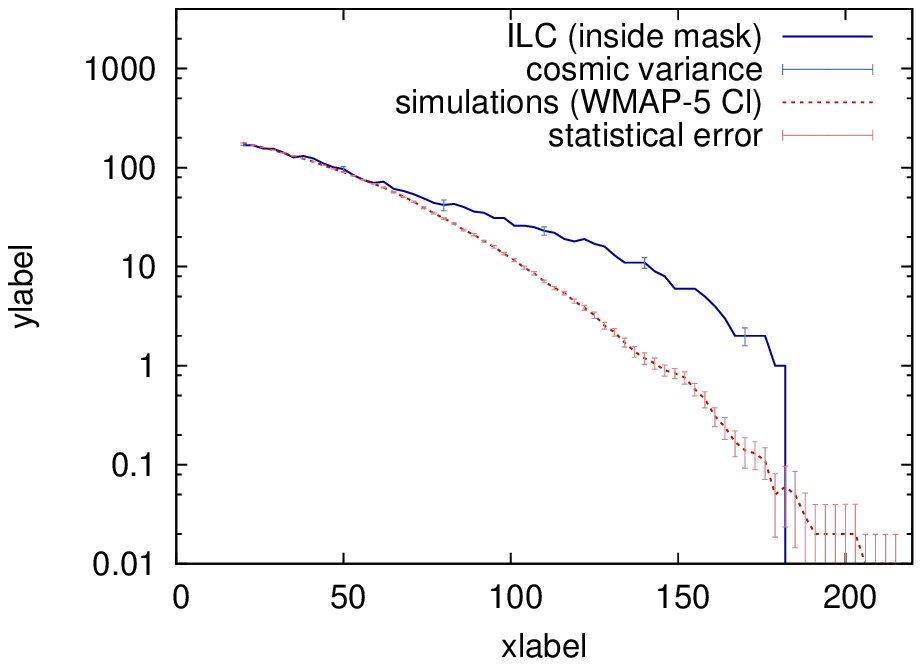}
		}
		\caption{Spot abundances in the ILC map (with cosmic variance) compared to
		simulations (with statistical errors) based on the measured
		\mbox{WMAP-5} $C_\ell$ on an
		angular scale of $a=6^\circ$ for three different parts of the
		sky. The corresponding values of $p_s$ are $p_s^\text{full} = 58\%$, $p_s^\text{out} = 7\%$, and $p_s^\text{in} =
		96\%$.}
		\label{fig:spots_ILC}
	\end{center}
\end{figure}

The five-year ILC map is the best approximate full-sky CMB map
available.  We therefore analyze it even though the quality of the
reconstruction is not high enough to guarantee robustness of the
results (see, also, Sec.~\ref{sec:data}). We analyze the ILC full-sky
map and 100 Gaussian full-sky simulations based on the \mbox{WMAP-5}
power spectrum and separately consider the results in three sky
regions.  First, we analyze the full sky. Second, we collect the
spots of those regions that have also been studied in the $V$ map,
i.e., regions with no or little overlap with the KQ75 mask (tolerant
selection). Finally, we consider the remaining spots that consequently
lie in sectors completely inside the mask or with considerable mask
overlap (rejected by tolerant selection). We loosely refer to the
three regions as {\it full sky}, {\it outside}, and {\it inside mask}.
The results are plotted in Fig.~\ref{fig:spots_ILC}.
In the previously analyzed region (outside the mask), we see too few
spots, as before. But there are by far too many spots in the
complementary region. The variances providing the error bars are, as
always, obtained from Eq.~(\ref{eq:errors_abundance}).  Although there
are less statistics inside the mask than in the full sky, the error
bars in the corresponding figure are smaller.  This can be intuitively
understood as follows. If, for simplicity, we assumed that the spot
abundances outside and inside the mask were statistically independent,
the variances $\sigma_{\text{in}}^2$, $\sigma_{\text{out}}^2$ would
add to $\sigma_\text{full}^2$ in the full sky, whence
$\sigma_{\text{in}} < \sigma_\text{full}$. The loss of statistics when
counting spots inside the masked region only causes the {\it relative}
fluctuations between two Gaussian simulations to increase.  The error
bars in the central figure (outside the mask) visually appear larger
due to the logarithmic plotting but are in fact smaller than for the
full sky.

The values of $p_s$ confirm the uneven distribution of spots in the
ILC map.  For the full sky, we have $p_s^\text{full} = 58\%$ in good
agreement with the simulations.  Outside the mask, there are too few
spots, $p_s^\text{out} = 7\%$, whereas inside the mask, we find
$p_s^\text{in} = 96\%$.  The ILC map is clearly anisotropic. Other
authors draw the same conclusion~\citep{Hajian07,Copi08,Bernui08}.

Anisotropy of the CMB is a possible explanation of the discrepancies
revealed in Sec.~\ref{sec:cut-sky} and quantified in
Table~\ref{tab:significance}, and indeed, the ILC map contains this
anisotropy. But since there is not enough reliable information about
the CMB signal in the galactic plane, we cannot finally judge whether
this is the true solution to the problem. We have also studied if,
additional to the galactic plane, the orientation of the galactic
halo defines a preferred direction. Therefore, we divided the ILC map
into two halves, one around the galactic center and one covering the
opposite direction. We have seen no signal of anisotropy in this
direction.

\subsection{Modified power spectra}\label{sec:modified}

We have pointed out that anisotropy is a potential explanation. It is
however unsatisfying to assume that so many additional spots lie in
the contaminated regions hidden by the KQ75 mask. This would be a
surprising coincidence of CMB signals and the orientation of the
galactic plane. Alternatively, we may stick to statistical isotropy;
then, our results may be due to some non-Gaussian signal. 

In this section, we investigate whether our results imply non-Gaussianity or
statistical anisotropy by themselves. We do this by analyzing the
effect of modifications to the $C_\ell$ spectrum.

\begin{figure}[htb!]
	\begin{center}
		\subfigure{
		\psfrag{xlabel}[B][c][.8][0]{Threshold
		$\Delta\mathfrak T$ [$\upmu \text{K}$]}
		\psfrag{ylabel}[B][c][.8][0]{Abundances of hot spots} 
		\includegraphics[width=.45\textwidth]{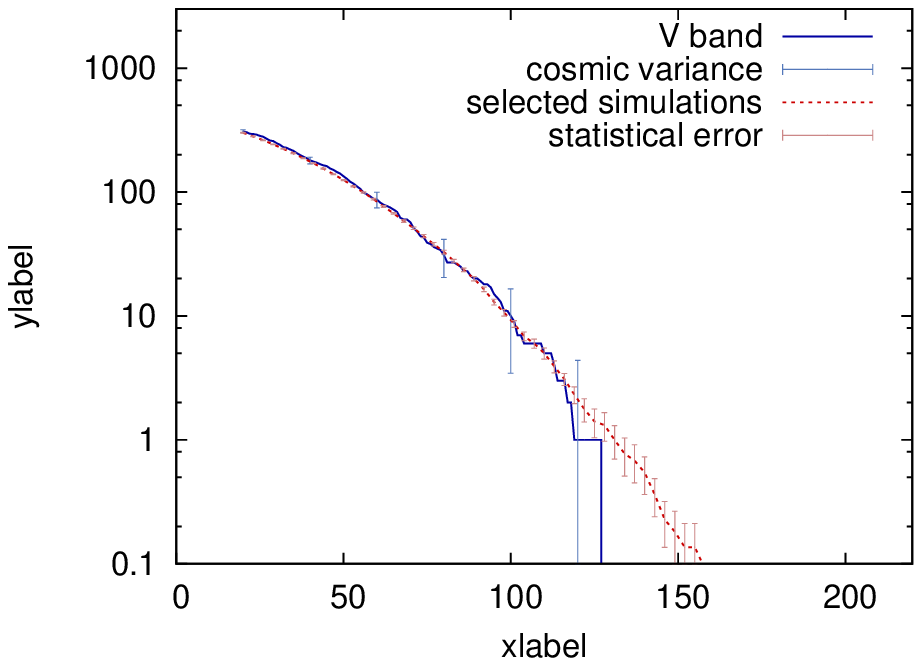}
		}
		\subfigure{
		\psfrag{xlabel}[B][c][.8][0]{Threshold
		$\Delta\mathfrak T$ [$\upmu \text{K}$]}
		\psfrag{ylabel}[B][c][.8][0]{Abundances of cold spots} 
		\includegraphics[width=.45\textwidth]{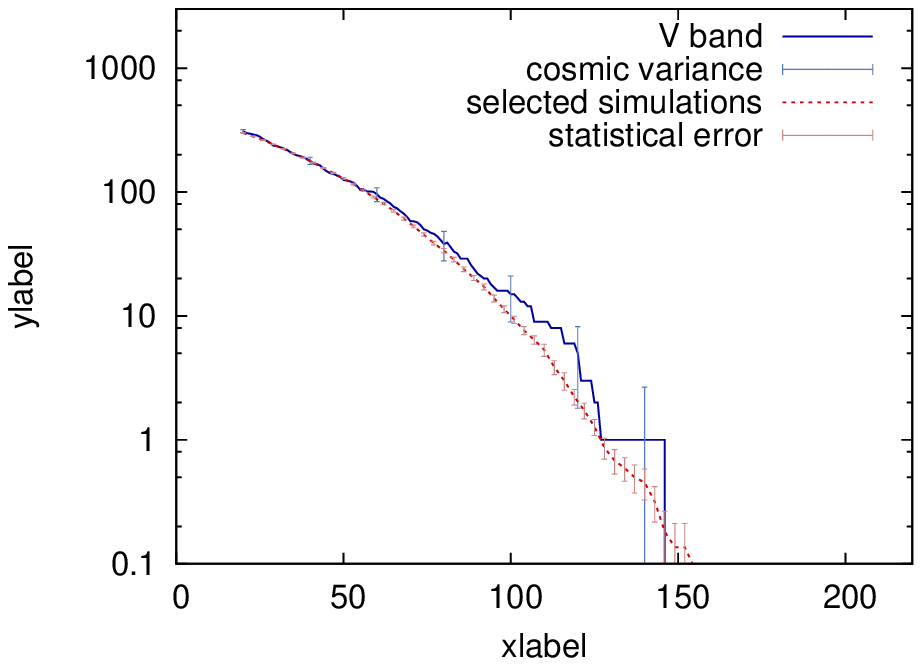}
		}
		\caption{Spot abundances of Gaussian simulations $k$ (errors statistical) with
		$\Delta T_{rms}^{(k)} \le \Delta T_{rms}^{\text{V
		map}}$ in comparison to the $V$ map (with cosmic variance).
		For these plots, we have $p_s^\text{hot} = 68\%$ and
		$p_s^\text{cold} = 86\%$, showing agreement. Since we only
		consider Gaussian simulations with $\Delta T_{rms}^{(k)}$
		{\it smaller} than in the $V$ map, it is no surprise that
		the $p_s$ values lie above $50\%$.}
		\label{fig:conform}
	\end{center}
\end{figure}

So far, $\Delta T_{rms}$ has proved to be a good parameter to quantify
the visible effects. We can perform a quick check whether our data
supports the hypothesis that $\Delta T_{rms}$ is the decisive
parameter. Out of the 500 simulations with the \mbox{WMAP-5} power
spectrum used in Sec.~\ref{sec:cut-sky}, we collect those with a
$\Delta T_{rms}$ smaller or equal than those found in the $V$ map.
Figure~\ref{fig:conform} shows their spot abundances which agree well
with the $V$ map.

\begin{figure}[htb!]
	\begin{center}
		\subfigure{
		\psfrag{xlabel}[B][c][.8][0]{Angular scale $a$
		[$^\circ$]}
		\psfrag{ylabel}[B][c][.8][0]{$\Delta T_{rms}$ [$\upmu
		\text{K}$]} 
		\includegraphics[width=.4\textwidth]{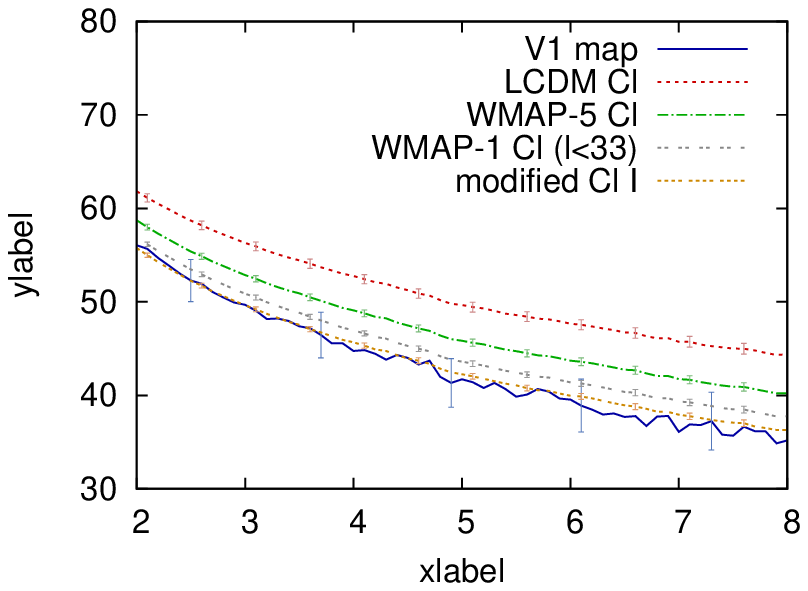}
		}
		\subfigure{
		\psfrag{xlabel}[B][c][.8][0]{Angular scale $a$
		[$^\circ$]}
		\psfrag{ylabel}[B][c][.8][0]{$\Delta T_{rms}$ [$\upmu
		\text{K}$]} 
		\includegraphics[width=.4\textwidth]{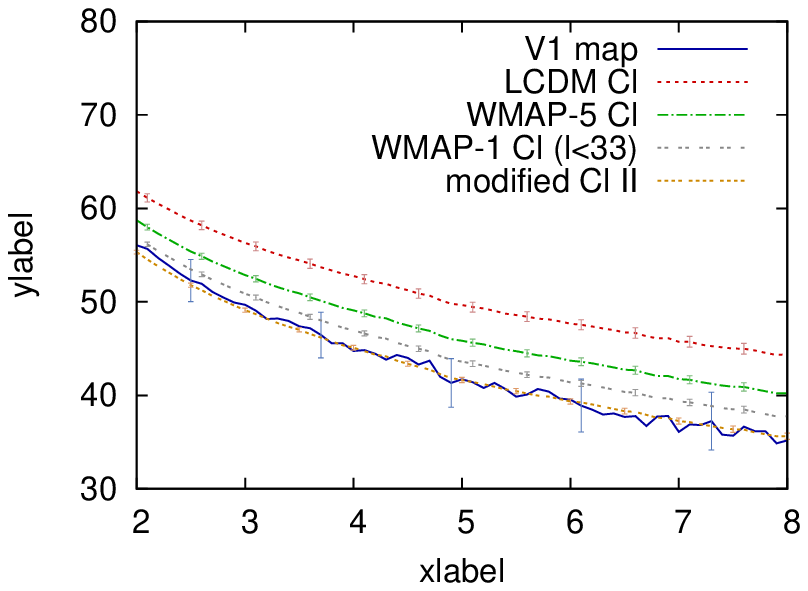}
		}
		\caption{The mean temperature fluctuation for large angular
		scales $a$. We compare the $V$1 map with $\Lambda$CDM
		simulations (highest $\Delta T_{rms}$), simulations based on
		the \mbox{WMAP-5} $C_\ell$ (a bit lower), the \mbox{WMAP-1} $C_\ell$ for $\ell<33$
		(still lower), and on two modified spectra. The first
		modification is created by setting $C_\ell=0$ for $\ell\le 3$, the
		second by halving the $C_\ell$ for $\ell \le 5$. The modified
		spectra agree well with the $V$1 map.}
		\label{fig:modified}
	\end{center}
\end{figure}

\begin{figure*}[htb]
	\begin{center}
		\psfrag{xlabel}[B][c][.8][0]{Threshold
		$\Delta\mathfrak T$ [$\upmu \text{K}$]}
		\psfrag{ylabel}[B][c][.8][0]{Abundances of hot spots} 
		\includegraphics[width=.45\textwidth]{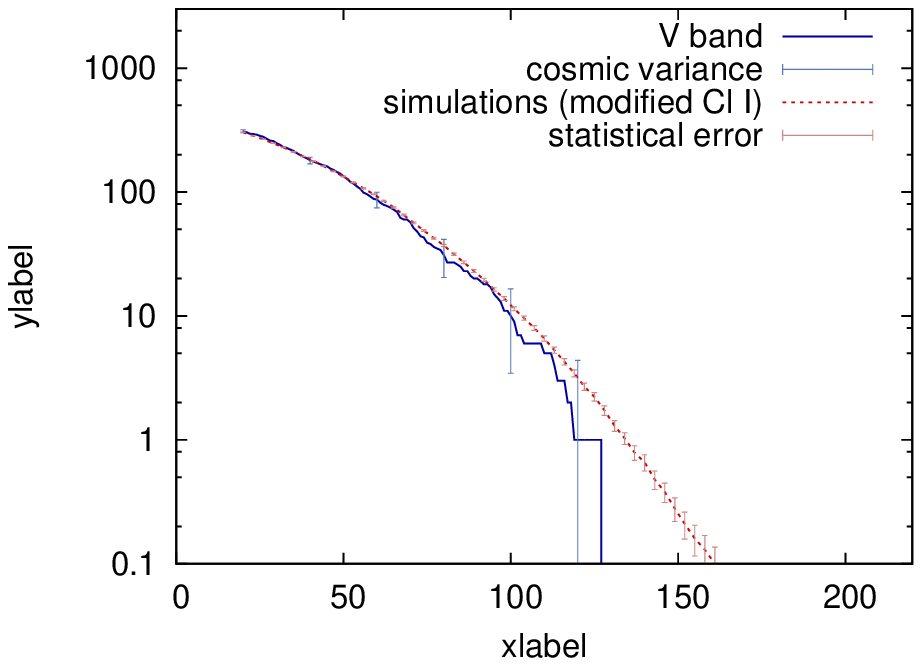}
		\psfrag{xlabel}[B][c][.8][0]{Threshold
		$\Delta\mathfrak T$ [$\upmu \text{K}$]}
		\psfrag{ylabel}[B][c][.8][0]{Abundances of hot spots} 
		\includegraphics[width=.45\textwidth]{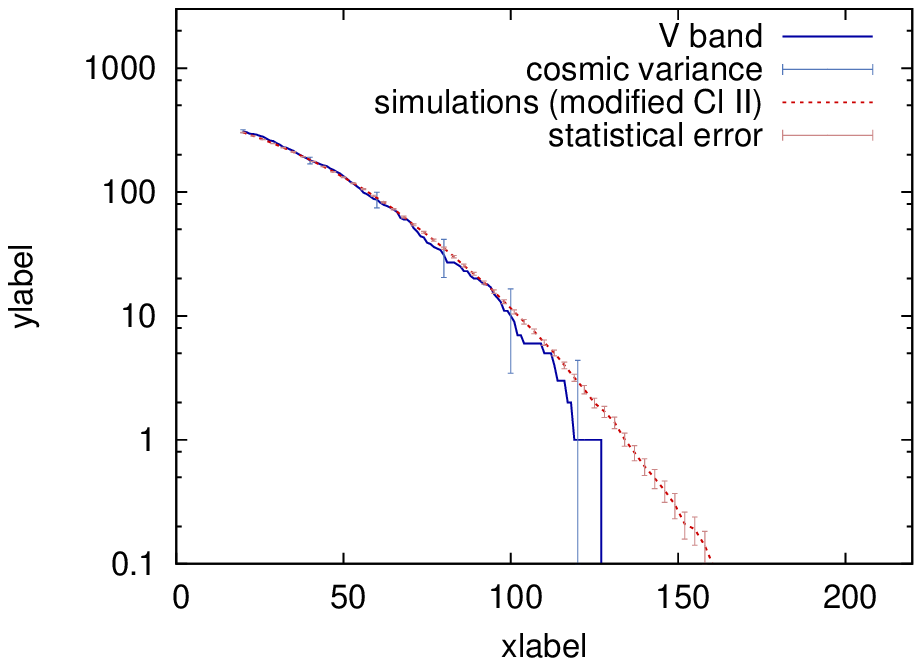}
		\psfrag{xlabel}[B][c][.8][0]{Threshold
		$\Delta\mathfrak T$ [$\upmu \text{K}$]}
		\psfrag{ylabel}[B][c][.8][0]{Abundances of cold spots}
		\includegraphics[width=.45\textwidth]{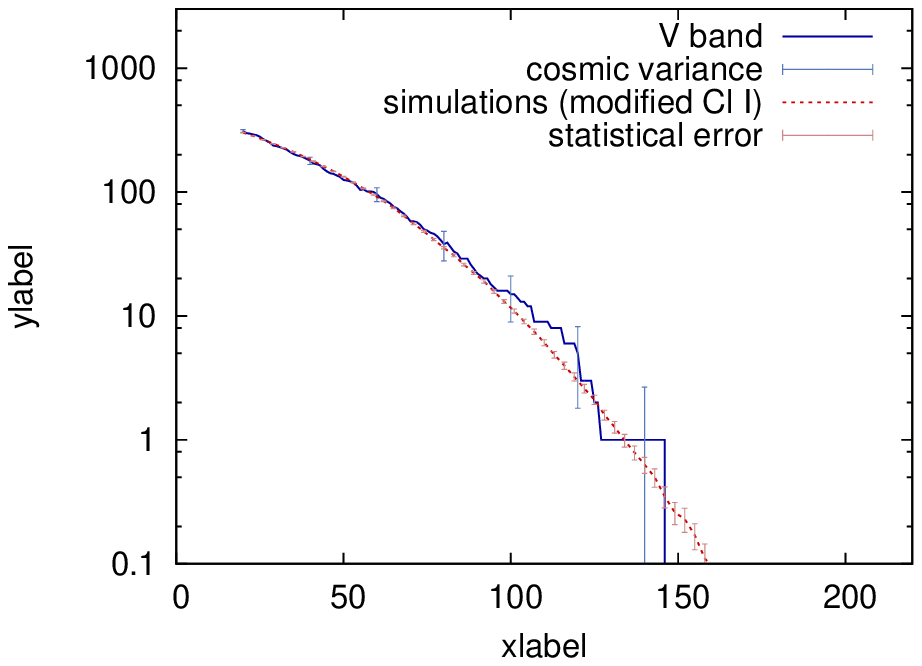}
		\psfrag{xlabel}[B][c][.8][0]{Threshold
		$\Delta\mathfrak T$ [$\upmu \text{K}$]}
		\psfrag{ylabel}[B][c][.8][0]{Abundances of cold spots}
		\includegraphics[width=.45\textwidth]{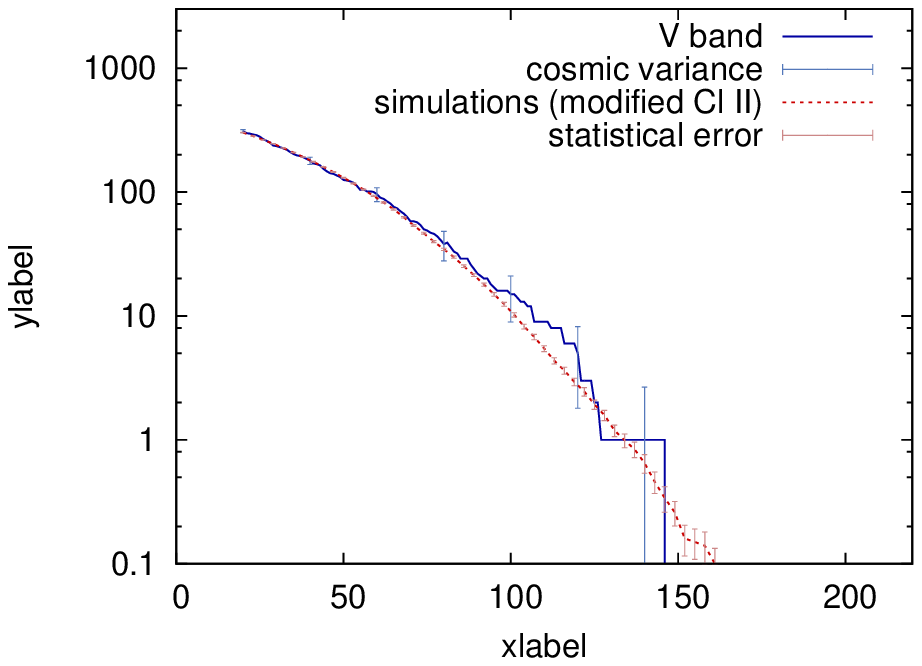}
		\caption{Spot abundances in the CMB sky (with cosmic variance) as compared to 100
		simulations based on the two modified spectra, respectively, (errors
	 statistical) on an angular scale of $a=6^\circ$. The first modification
		yields $p_s^\text{hot} = 30\%$ and $p_s^\text{cold}=55\%$, the
		second modification $p_s^\text{hot} = 37\%$ and
		$p_s^\text{cold} = 67\%$.}
		\label{fig:spots_V_modified}
	\end{center}
\end{figure*}

If there is a $C_\ell$ spectrum that produces $\Delta T_{rms}$ values similar
to the ones found in the $V$ map, our results alone do not imply
non-Gaussianity or statistical anisotropy.
In order to keep the analysis as generic as possible, we
do not use any specific cosmological model but only modify the $C_\ell$
of the original \mbox{WMAP-5} spectrum. Figure~\ref{fig:sigmas_original}
suggests that only large scales are affected which is why we
concentrate on a few low multipoles $\ell$. \citet{Copi08} found that the
correlation function is essentially zero on angular scales above
$\approx 60^\circ$. Since this scale is roughly linked to the
multipole range $\ell \le 3$, our first modification simply consists in
setting $C_\ell\equiv 0$ for $\ell \le 3$ (although of course the
correlation function does not translate this easily).  Another example
may be to halve the $C_\ell$ for $\ell \le 5$ (modification II).
Figure~\ref{fig:modified} shows the resulting values of $\Delta
T_{rms}$.
The plots show the discrepancies between the $\Lambda$CDM prediction,
the \mbox{WMAP-5} spectrum, and observation. We also show the results for a
combined power spectrum, replacing the first $32$ multipoles by the
values quoted by \mbox{WMAP-1}~\citep{Hinshaw03}. For this range of
multipoles the WMAP analysis changed after the 1-year release,
following the suggestion of \citet{Efstathiou03}. The difference
between \mbox{WMAP-5} and \mbox{WMAP-1} may serve as an illustration that the
extraction of reliable $C_\ell$ for low $\ell$ is a nontrivial task. 
Modifications I and II of the power spectrum succeeded in reconciling
Gaussian simulations and observed CMB sky. This is confirmed by
measuring the spot abundances in simulated maps based on the modified
spectra, seen in Fig.~\ref{fig:spots_V_modified}.
We conclude that our results are not incompatible with Gaussianity.
However, if we stick to Gaussianity, they favor (although
statistically not significant, cf.
Table~\ref{tab:significance_measured}) even lower values of the first
multipoles than currently estimated.

\section{Conclusions}\label{sec:discussion}

The study of spot abundances has revealed discrepancies between the
cut-sky CMB temperature maps and the standard best-fit $\Lambda$CDM
model or, but less significant, a Gaussian spectrum for the $C_\ell$
estimated by \mbox{WMAP-5}. We have shown in Sec.~\ref{sec:modified}
that a good parameter to quantify them is the mean temperature
fluctuation $\Delta T_{rms}$ which we investigated on large scales. On
scales $a$ between $4^\circ$ and $8^\circ$, only $0.16\%$ to $0.62\%$
of Gaussian simulations based on the $\Lambda$CDM best-fit power
spectrum fall below the $\Delta T_{rms}$ value of the observed $V$
map. If this merely was an imprint of the anomalously low
quadrupole, we would expect the discrepancies to disappear when
removing the quadrupole from the Gaussian simulations and the $V$
map. The difference in fact reduces, the aforementioned fractions
change to $2.5\%$ to $8.0\%$. These numbers are not significant and do
not allow for a clear interpretation whether our results go beyond the
quadrupole anomaly. Similar fractions are obtained when exchanging
the $\Lambda$CDM best-fit spectrum by the originally published
\mbox{WMAP-5} $C_\ell$, yielding $2.4\%$ to $5.8\%$. This is difficult
to understand if we bear in mind that the $C_\ell$ themselves are
estimated from the cut-sky CMB maps~\citep{Nolta09}.

Non-Gaussianity and also statistical anisotropy are possible
explanations. In our case, anisotropy means that many spots have to be
hidden behind the masked region. Unfortunately, this hypothesis can
hardly be tested as there is currently no method to reliably extract
the CMB signal in the highly foreground-contaminated regions.
Nonetheless, we have employed the \mbox{WMAP-5} ILC full-sky CMB map
and found evidence for anisotropy in this map. This agrees with
results obtained by \citet{Hajian07} and \citet{Copi08} who found that
most of the power on the largest scales comes from the (masked) galaxy
region. Though possible, this unnatural alignment of the CMB signal
with the galactic plane would be intriguing and lacks so far any
explanation.

Our analysis of Sec.~\ref{sec:modified} shows that our results for
cut-sky maps do not suggest non-Gaussianity or statistical anisotropy
by themselves. They agree well with Gaussian fluctuations if one
performs a modification of the lowest multipoles. In doing so, no
fine-tuning of the $C_\ell$ is necessary in order to reconcile the spot
abundances from Gaussian simulations and the observed CMB. It is
sufficient to lower the first multipoles by a substantial amount. When
studying local extrema in the temperature field, \citet{Hou09}
similarly found discrepancies that disappeared when excluding the
first multipoles. We recall, however, that the $C_\ell$ and the
assumption of Gaussianity completely fix the expected spot abundances.
If both the extraction of the $C_\ell$ by \mbox{WMAP-5} and our analysis
of spot abundances are correct, our results may indicate
non-Gaussianity or statistical anisotropy.

If the discrepancies are not caused by mere statistical coincidence or
unknown secondary effects, we have to leave open the question whether
we see the consequence of non-Gaussianity or anisotropy, or whether
our results strengthen the evidence for a severe lack of large-scale
power. The first case would challenge fundamental assumptions, the
second would make it difficult to understand the CMB maps on large
scales within standard $\Lambda$CDM cosmology. If the discrepancies
between the $C_\ell$, as determined by \mbox{WMAP-5}, and the spot
abundances persist, this can be interpreted as a signal for
non-Gaussian fluctuations.

\section*{Acknowledgements}

We would like to thank Christian T. Byrnes for useful discussions. We
also thank the WMAP team for producing great data products and
publishing them on LAMBDA (the Legacy Archive for Microwave Background
Data Analysis). Support for LAMBDA is provided by the NASA Office of
Space Science. We acknowledge the use of the HEALPix
package~\citep{HEALPIX} that we employed for many tasks, most notably
the creation and preparation of Gaussian simulations.

\bibliography{cmbspots}

\begin{thebibliography}{}

\bibitem[Bernui and Reboucas, 2009]{Bernui08}
Bernui, A. and Reboucas, M.~J. (2009).
\newblock {Searching for non-Gaussianity in the WMAP data}.
\newblock {\em Phys. Rev.}, D79:063528, arXiv:0806.3758.

\bibitem[Bernui et~al., 2006]{Bernui06}
Bernui, A., Villela, T., Wuensche, C.~A., Leonardi, R., and Ferreira, I.
  (2006).
\newblock {On the CMB large-scales angular correlations}.
\newblock {\em Astron. Astrophys.}, 454:409--414, arXiv:astro-ph/0601593.

\bibitem[Cabella et~al., 2004]{Cabella04}
Cabella, P., Hansen, F., Marinucci, D., Pagano, D., and Vittorio, N. (2004).
\newblock {Search for non-Gaussianity in pixel, harmonic and wavelet space:
  compared and combined}.
\newblock {\em Phys. Rev.}, D69:063007, arXiv:astro-ph/0401307.

\bibitem[Copi et~al., 2009]{Copi08}
Copi, C.~J., Huterer, D., Schwarz, D.~J., and Starkman, G.~D. (2009).
\newblock {No large-angle correlations on the non-Galactic microwave sky}.
\newblock {\em Mon. Not. Roy. Astron. Soc.}, 399:295--303, arXiv:0808.3767.

\bibitem[de~Oliveira-Costa et~al., 2004]{deOliveiraCosta03}
de~Oliveira-Costa, A., Tegmark, M., Zaldarriaga, M., and Hamilton, A. (2004).
\newblock {The significance of the largest scale CMB fluctuations in WMAP}.
\newblock {\em Phys. Rev.}, D69:063516, arXiv:astro-ph/0307282.

\bibitem[Durrer, 2008]{Durrer08}
Durrer, R. (2008).
\newblock {\em {The Cosmic Microwave Background}}.
\newblock Cambridge University Press.

\bibitem[Efstathiou, 2004]{Efstathiou03}
Efstathiou, G. (2004).
\newblock {A Maximum Likelihood Analysis of the Low CMB Multipoles from WMAP}.
\newblock {\em Mon. Not. Roy. Astron. Soc.}, 348:885, arXiv:astro-ph/0310207.

\bibitem[Eriksen et~al., 2004]{Eriksen03}
Eriksen, H.~K., Hansen, F.~K., Banday, A.~J., Gorski, K.~M., and Lilje, P.~B.
  (2004).
\newblock {Asymmetries in the CMB anisotropy field}.
\newblock {\em Astrophys. J.}, 605:14--20, arXiv:astro-ph/0307507.

\bibitem[Gold et~al., 2009]{Gold09}
Gold, B. et~al. (2009).
\newblock {Five-Year Wilkinson Microwave Anisotropy Probe (WMAP) Observations:
  Galactic Foreground Emission}.
\newblock {\em Astrophys. J. Suppl.}, 180:265--282, arXiv:0803.0715.

\bibitem[Gorski et~al., 2005]{HEALPIX}
Gorski, K.~M. et~al. (2005).
\newblock {HEALPix -- a Framework for High Resolution Discretization, and Fast
  Analysis of Data Distributed on the Sphere}.
\newblock {\em Astrophys. J.}, 622:759--771, arXiv:astro-ph/0409513.

\bibitem[Hajian, 2007]{Hajian07}
Hajian, A. (2007).
\newblock {Analysis of the apparent lack of power in the cosmic microwave
  background anisotropy at large angular scales}.
\newblock arXiv:astro-ph/0702723.

\bibitem[Hansen et~al., 2009]{Hansen08}
Hansen, F.~K., Banday, A.~J., Gorski, K.~M., Eriksen, H.~K., and Lilje, P.~B.
  (2009).
\newblock {Power Asymmetry in Cosmic Microwave Background Fluctuations from
  Full Sky to Sub-degree Scales: Is the Universe Isotropic?}
\newblock {\em Astrophys. J.}, 704:1448--1458, arXiv:0812.3795.

\bibitem[Hill et~al., 2009]{Hill09}
Hill, R.~S. et~al. (2009).
\newblock {Five-Year Wilkinson Microwave Anisotropy Probe (WMAP) Observations:
  Beam Maps and Window Functions}.
\newblock {\em Astrophys. J. Suppl.}, 180:246--264, arXiv:0803.0570.

\bibitem[Hinshaw et~al., 2003]{Hinshaw03}
Hinshaw, G. et~al. (2003).
\newblock {First Year Wilkinson Microwave Anisotropy Probe (WMAP) Observations:
  Angular Power Spectrum}.
\newblock {\em Astrophys. J. Suppl.}, 148:135, arXiv:astro-ph/0302217.

\bibitem[Hinshaw et~al., 2007]{Hinshaw07}
Hinshaw, G. et~al. (2007).
\newblock {Three-year Wilkinson Microwave Anisotropy Probe (WMAP) observations:
  Temperature analysis}.
\newblock {\em Astrophys. J. Suppl.}, 170:288, arXiv:astro-ph/0603451.

\bibitem[Hoftuft et~al., 2009]{Hoftuft09}
Hoftuft, J. et~al. (2009).
\newblock {Increasing evidence for hemispherical power asymmetry in the
  five-year WMAP data}.
\newblock {\em Astrophys. J.}, 699:985--989, arXiv:0903.1229.

\bibitem[Hou et~al., 2009]{Hou09}
Hou, Z., Banday, A.~J., and Gorski, K.~M. (2009).
\newblock {The Hot and Cold Spots in Five-Year WMAP Data}.
\newblock arXiv:0903.4446.

\bibitem[Land and Magueijo, 2005]{Land05}
Land, K. and Magueijo, J. (2005).
\newblock {The axis of evil}.
\newblock {\em Phys. Rev. Lett.}, 95:071301, arXiv:astro-ph/0502237.

\bibitem[Larson and Wandelt, 2004]{Larson04}
Larson, D.~L. and Wandelt, B.~D. (2004).
\newblock {The Hot and Cold Spots in the WMAP Data are Not Hot and Cold
  Enough}.
\newblock {\em Astrophys. J.}, 613:L85--L88, arXiv:astro-ph/0404037.

\bibitem[Larson and Wandelt, 2005]{Larson05}
Larson, D.~L. and Wandelt, B.~D. (2005).
\newblock {A Statistically Robust 3-Sigma Detection of Non- Gaussianity in the
  WMAP Data Using Hot and Cold Spots}.
\newblock arXiv:astro-ph/0505046.

\bibitem[McEwen et~al., 2008]{McEwen08}
McEwen, J.~D., Hobson, M.~P., Lasenby, A.~N., and Mortlock, D.~J. (2008).
\newblock {A high-significance detection of non-Gaussianity in the WMAP 5-year
  data using directional spherical wavelets}.
\newblock arXiv:0803.2157.

\bibitem[Monteserin et~al., 2008]{Monteserin07}
Monteserin, C. et~al. (2008).
\newblock {A low CMB variance in the WMAP data}.
\newblock {\em Mon. Not. Roy. Astron. Soc.}, 387:209--219, arXiv:0706.4289.

\bibitem[Nolta et~al., 2009]{Nolta09}
Nolta, M.~R. et~al. (2009).
\newblock {Five-Year Wilkinson Microwave Anisotropy Probe (WMAP) Observations:
  Angular Power Spectra}.
\newblock {\em Astrophys. J. Suppl.}, 180:296--305, arXiv:0803.0593.

\bibitem[Page et~al., 2003]{Page03}
Page, L. et~al. (2003).
\newblock {The Optical Design and Characterization of the Microwave Anisotropy
  Probe}.
\newblock {\em Astrophys. J.}, 585:566--586, arXiv:astro-ph/0301160.

\bibitem[Vielva et~al., 2004]{Vielva03}
Vielva, P., Martinez-Gonzalez, E., Barreiro, R.~B., Sanz, J.~L., and Cayon, L.
  (2004).
\newblock {Detection of non-Gaussianity in the WMAP 1-year data using spherical
  wavelets}.
\newblock {\em Astrophys. J.}, 609:22--34, arXiv:astro-ph/0310273.

\bibitem[Yadav and Wandelt, 2008]{Yadav07}
Yadav, A. P.~S. and Wandelt, B.~D. (2008).
\newblock {Evidence of Primordial Non-Gaussianity $(f_{\rm NL})$ in the
  Wilkinson Microwave Anisotropy Probe 3-Year Data at 2.8$\sigma$}.
\newblock {\em Phys. Rev. Lett.}, 100:181301, arXiv:0712.1148.

\end{thebibliography}

\end{document}